%% file: hierarchie.tex
\documentclass[a4paper]{article}

\usepackage{amssymb,amscd}
\usepackage{amsmath,amsfonts,latexsym}
\usepackage[dvips]{graphicx}
\usepackage{color}
\usepackage[latin1]{inputenc}
\usepackage{enumerate,afterpage}
\usepackage{tikz}
\usetikzlibrary{matrix,arrows}
\usepackage[all]{xy}
\usepackage{url}

\title{The prestige and status of research fields within mathematics}

\author{Jean-Marc Schlenker$^{1\ast}$ \\
  \normalsize{$^{1}$Department of mathematics, University of Luxembourg} \\
   \normalsize{$^\ast$To whom correspondence should be addressed; E-mail: jean-marc.schlenker@uni.lu}}
\date{v1, \today}
 
\begin{document}


\newcounter{notes}%
\newcommand{\marginnote}[1]{
\refstepcounter{notes}  
\nolinebreak
$\hspace{-5pt}{}^{\text{\tiny \rm \arabic{notes}}}$
\marginpar{\tiny \arabic{notes}) #1}
}

\maketitle

\begin{abstract}
  While the ``hierarchy of science'' has been widely analysed, there is no corresponding study of the status of subfields within a given scientific field. We use bibliometric data to show that subfields of mathematics have a different ``standing'' within the mathematics community. Highly ranked departments tend to specialize in some subfields more than in others, and the same subfields are also over-represented in the most selective mathematics journals or among recipients of top prizes. Moreover this status of subfields evolves markedly over the period of observation (1984--2016), with some subfields gaining and others losing in standing. The status of subfields is related to different publishing habits, but some of those differences are opposite to those observed when considering the hierarchy of scientific fields.

  We examine possible explanations for the ``status'' of different subfields. Some natural explanations -- availability of funding, importance of applications -- do not appear to function, suggesting that factors internal to the discipline are at work. We propose a different type of explanation, based on a notion of ``focus'' of a subfield, that might or might not be specific to mathematics.
\end{abstract}

\tableofcontents

\section{Motivations, questions and results}


\subsection{The prestige of subfields, and why they matter}

Certain fields of science benefit from a higher degree of  prestige and status than others. This ``status'' of fields or subfields can vary in time and between country, but it has deep consequences on the advancement of science, for instance:
\begin{itemize}
\item Through the allocation of talented researchers, since talented students tend to disproportely choose some fields over others.
\item Through its impact on institutional policy, and the choice by universities or research institutions to invest in some fields more than others, depending on their profile and objectives.
\item Through its impact on science policy, and the allocation by governement or research funding agencies of priorities in research.
\end{itemize}

The prestige of a field should be distinguished from how {\em fashionable} it is. Some fields sometimes undergo rapid progress and promise to have a huge impact in the short or medium term (a recent example is for instance found in deep learning) and therefore attract both researchers and funding. Other fields can attract talented students because they are considered as stepping stones towards successful careers. We will see that, in mathematics, a clear distinction should be made between fields which are considered as ``high status'' and those offering the best career perspective -- in fact, a case could be made that there is a {\em negative} relation between the two notions, see Section \ref{ssc:explanations}.

We investigate whether and why some directions of research appear to be valued by top departments or top journals more than others, using quantitative and bibliometric tools and methods. It is difficult to compare different fields, since they tend to have different publishing practices, venues, etc. It is easier, at least when using bibliometric tools, to consider differences between subfields of a given field. Here we focus on mathematics, and on the dynamics of its different subfields. 

The analysis of status between subfields of mathematics should be compared to the wider debate and ample bibliography on a ``hierarchy of sciences'' going back at least to Auguste Comte, see e.g. \cite{comte1835cours,cole:hierarchy}. The main indicator of a higher ``status'' of a field is generally considered to be the {\em consensus level}, which is also related to the level of {\em complexity} of the object of study. The reader can find in \cite{fanelli2013} an analysis of the relations between the hierarchy/level of consensus of a field and different types of bibliometric characteristics, such as the average number of authors, number of pages, or number of cited references. For instance, it appears that at the level of fields, shorter papers or less old references should be related to higher consensus level and ``status'', which might contrast with the situation within subfields of mathematics (see below).

Among scientific fields, mathematics is generally considered as having a high level of consensus. In a slightly different direction of analysis, proposed by Hargens \cite{hargens1975}, see also \cite{braxton1996variation}, mathematics is characterized by a high level of {\em normative integration} -- the sharing of common beliefs and values -- but a low level of {\em functional integration} -- the activity of mathematicians does not depend directly much on those of her colleagues. 


\bigskip

\subsection{Questions, observations, results}

Mathematicians often have strong preferences for some fields over others, and preferences vary widely. However they also often express the idea that some fields of mathematics are ``more difficult'', ``more central'', or ``more important'' than others, and some convergence seems to appear in those opinions. While the assessment of each subfield varies from one author to the other, it is interesting to ask whether there is a ``general pattern'' in the way different mathematicians gauge the ``difficulty'', ``centrality'' or ``importance'' of a each field, and to explain it. One might expect that the ``importance'' attributed to subfields is directly related to their relevance to applications in science and technology -- we will see however that this does not seem to be the case. 

This observation leads to several questions.
\begin{itemize}
\item Can a ``hierarchy'' of mathematical fields be observed, and how?
\item Do different observations provide similar rankings of fields?
\item Is this ranking stable over time? If not, how does it evolve?
\item How can it be explained?
\end{itemize}

One way to approach those questions is through existing and established hierarchies:
\begin{itemize}
\item of mathematics journals (as measured here by their MCQ, a kind of IF adapted to mathematics),
\item of mathematics departments. 
\end{itemize}
We can assume that if some fields were more valued than others, the ``best'' journals and the ``best'' departments would focus more on those fields than on others, while less prestigious departments and journals would have to do with less prestigious fields. Those observations, confirmed by a set of bibliometric observations, will serve as the basis for the analysis conducted here.

Some of the key findings presented below are:
\begin{enumerate}
\item There is a clear ``ranking'' of subfields of mathematics, with some more prestigious fields having a much larger relative share of the mathematics departments of the most prestigious universities, of the papers published in the most respected journals (Section \ref{ssc:top3}), of the top prizes (Section \ref{ssc:fields}).
\item Different ``measures'' of the rank of fields within mathematics tend to give very similar results, compare Section 3, 4 and 5.
\item The subfields with the highest apparent ``status'' are typically the most abstract ones and those with little direct applications, such as Algebraic Geometry, while applied fields like numerical modelling or statistics have low status -- this in spite of efforts by some funding agencies to promote precisely those low-status fields.
\item The ``status'' of subfields can certainly not be explained by commonly used notions of impact, such as citation count, since the most prestigious fields tend to be those with lowest citations numbers, and conversely. 
\item The prestige of some fields has changed quite dramatically over the period of observation considered here (1984--2016) in one direction or in the other. For instance the prestige of Differential geometry or of Analysis seems to have declined considerably, while those of Probability or Partial Differential Equations has increased markedly.
\item The status of a subfield is related to differences in publishing habits: higher status is rather strongly related to less authors/article, and to longer articles.
\item The recruitment strategy of departments varies greatly with their ``status'', measured here by their ranking according to an indicator of production. Top departments tend to favor disproportionately the ``noble'' subfields, while low-ranking departments hire more experts of subfields with a lower status, see Section \ref{ssc:recruitments}.
\item There is a significant mobility of mathematicians between subfields. This mobility is higher in highly ranked universities, and tends to follow the evolution of the ``status'' of fields -- mathematicians at better departments tend to move out of subfields of decreasing status, and into subfields of improving status, see Section \ref{ssc:mobility}.
\item The scientific productivity of authors is different between subfields. In some high-status fields it tends to be highly concentrated on a small number of highly productive authors, while in other fields it decreases more slowly, see Section \ref{ssc:differing}. This fact might explain why the optimal strategy of different departments might to focus on some fields more than others, see Section \ref{ssc:strategies}.
\end{enumerate}

Along the way, we consider in Section \ref{ssc:indicators} what indicator of scientific productivity should be used for authors. We use data on grants from the European Research Council to calibrate different indicators of production based on the MCQ of journals (see Section \ref{sc:data} below) and come to the conclusion that indicators that best fit with the opinions of panels of mathematicians tend to give a much higher weight to a small number of journals with a high MCQ. 

Section \ref{sc:explanations} is dedicated to possible explanations of the ``value'' attributed to different fields. We consider several characteristics of subfields of mathematics that could be related to how ``prestigious'' an area of research is. We then concentrate on the 
  {\em focus} of a subfield: to what extent researchers in this subfield tend share an interest for a small number of questions (or conjectures) which therefore become important. The notion of focus considered here is related to, but distinct from, the notion of level of consensus in a field. It is closer to the notion of {\em normative integration} considered by Hargens \cite{hargens1975}, but with a twist: what is shared is not only beliefs and values, but more specifically the interest in a small set of questions or conjectures which are considered as particularly important.

  We will see that this notion can explain at least to some extend why some subfields of mathematics seem to enjoy a higher degree of prestige than others. Section \ref{ssc:focus} shows how a keyword analysis -- the use of the word ``conjecture'' in the title of articles or their {\em Math Reviews} entry -- correlates well with other measures of ``status'' identified here.

  \subsection{Why study mathematics?}
  
  Mathematics is a suitable field of study for the questions considered here for a number of reasons. It is a ``international'' scientific field, present in one form or another in almost all universities (since mathematics teaching is always present). It has well-defined fields and subfields, with relatively clear boundaries. The number of authors in each article is usually relatively low, allowing for better identification of author's roles. Another interesting feature is that the level of funding is probably less important for research in mathematics than in experimental fields, so that decision on topics to study might be less influenced by the availability of funding.

  Mathematics as a field has other significant feature that make it an interesting object of study.
  \begin{itemize}
  \item It is fully international, in the sense that while some nations tend to specialize more in some subfields, there is no main difference either in the main questions being considered or in the methods applied by mathematicians across the world. This differs from some fields (e.g. economics) where cultural and political factors can play a major role.
  \item Mathematics is a {\em large} field, measured by the number of active mathematicians. For instance, the number of mathematicians with a faculty position in US post-secondary institutions was estimated in 2017 as 25,632 (see  \cite{golbeck2019fall}), to be compared to a total of 822,513 across all disciplines \cite[Table 315.20]{snyder2019digest} (so that mathematicians represents close to 3\% of faculty in the US). By comparison, the number of physicists with a full-time faculty position in US post-secondary institution was estimated at 10100, so 2.5 times less than mathematicians.\footnote{A similar picture is obtained through the U.S. Bureau of Labor Statistics data on the somewhat larger group of ``Postsecondary teachers'', which includes professors at post-secondary institutions. In 2018 they estimated the total number at 1,350,700, including 51,250 in mathematics, 13,780 in physics, and 13,270 in economics. See \url{https://www.bls.gov/ooh/education-training-and-library/postsecondary-teachers.htm}} The size comparison between fields might vary from one country to the other, but one can expect that mathematics remains one of the larger academic fields.
  \item Mathematics is also an ancient field of study, with continuous development in the last centuries. As a consequence of this development, mathematics has branched into a variety of subfields, each with its own problems, methods and traditions. 
  \item Mathematics is also well determined as a field, with a clear definition: mathematicians insist on {\em proving} the statements that they consider as results. This distinguishes them quite clearly from scientists from other scientists, who might be interested in the same questions and the same objects but with different methods. This clear line can be seen for instance in mathematical physics, with on one side mathematicians intending to prove results, on the other physicists who do not need a formal proof once they have reached a high degree of confidence in the truth of a statement. Similarly, computational engineers might be satisfied by an program providing numerical approximations of solutions of a partial differential equations that are close to observed solutions, while applied mathematicians tend to search for a {\em proof} that the numerical solutions are close to the real ones.  
  \end{itemize}

Mathematics is also a convenient object of study thanks to the carefully curated and very complete bibliographic information that is available, in particular through {\em Mathematical Reviews}, a database produced by the {\em American Mathematical Society}. It has a number of very interesting features that don't have analogs in most other fields, for instance:
\begin{itemize}
\item It provides a clear identification of each institution and even each department within a given instition (each department is assigned a code, composed of a first block describing the country and a second block the institution).
\item It assigns to each individual author a number, so that even authors with the same first and last name can be clearly distinguished.
\item Each article is given a primary and one or several secondary {\em Mathematics Subject Classification} (M.S.C.) code, so that one can easily determine to what extend a given article is related to a certain subfield of mathematics.
\end{itemize}

We believe that a hierarchy of subfields might also be observed also in other fields of science. One can for instance find a hint in the description in \cite{nature:barry-simon} of the attitude of high-energy physicists towards condensed-matter physics. 

  \section{Data description}
  \label{sc:data}

  \subsection{A database of selected mathematics articles}
  
We use data collected selectively from {\em Math Reviews}, a journal published by the {\em American Mathematical Society} containing synopsis of most articles published in the area of mathematics as well as in some related fields. The data available through {\em Math Reviews} has a number of important qualities that make it particularly useful for bibliometric studies on mathematics. For instance, it attributes a unique code to each author -- even when two authors have the same first and last name. It also attributes to each articles a series of precises disciplinary classification codes. Finally the {\em Math Reviews} database attributes stable codes to each institution, so that affiliations of authors can be easily followed.

Within the {\em Math Reviews} database, we selected a list of approximatively 140 journals, which can be considered to be leading journals in pure and applied mathematics. The choice of the journals was made in two steps:
\begin{itemize}
\item A first list of journals was selected for a previous study \cite{drs}. At the time the journals selected were those with the highest impact factor among those having a cited half-life above a threshold, according to the {\em Journal Citation Report} 2006. 
\item More recently, this list was completed by adding the journals ranked as $A^*$ in the journal list of the {\em Australian Math. Society}
  \footnote{See \url{https://www.austms.org.au/Rankings/AustMS_final_ranked.html}}
  for mathematical sciences, one of the very few journal rankings in mathematics produced by consulting with experts (rather than using only bibliometric data). This list can be considered as a reasonable proxy for a ``consensus'' among mathematicians of which journals are most selective and important. 
\end{itemize}

For each journal, we used the full list of articles published between 1984 and 2016, recording for each article the most relevant data only (journal name, publication year, number of pages, number of authors, MSC (Mathematics Subject Classification) codes, and for each author, the MR code of the author and his/her affiliation). The total number of articles in the database that we used is 247 677. We assume that although this list of papers represents only a small fraction of the whole mathematics literature published during the period of study, it does contain most of the articles considered as really significant by mathematicians.

The list of journals considered here can be considered as somewhat arbitrary -- while most of those journals should be in any study intending to study the most relevant papers in mathematics, some of the less central journals could be replaced by others with a similar standing. Moreover this collection of journals gives a certain weight for each subfield, and different choices of journals would lead to different weighting of the subfields. However this weight of each field is not so important for the considerations made here, since our analysis of the status of subfields is based on the {\em differences} in weights of subfields in different departments (resp. journals, etc) rather than on the weight themselves. 

\subsection{A database of authors and PhDs}

The data from {\em Mathematical Reviews} was then merged with data from {\em Mathematical Genealogy}, a remarkably complete and apparently quite accurate freely accessible database of mathematics PhD thesis. It appears that most ``active'' mathematicians appearing in our data from {\em Mathematical Reviews} (those with at least 2 papers) also appear in {\em Mathematical Genealogy}. This second source contains additional information which is used in some of the results presented below, including the date of the defense and the PhD-granting institution for many PhD theses in mathematics.

Most of the data presented below are followed over the time frame of the study, from 1984 to 2016. To avoid too much random noise, we consider 8 periods each of 4 years (except the first one, 1984-1988, which is 5 years long).

\subsection{A limited list of subfields of mathematics}

For the purpose of this article, we defined a small group of subfields of mathematics, each corresponding to a small set of 2-digit MSC codes:
\begin{itemize}
\item {\em Algebra}, corresponding to codes 06, 08, 20, 18, 15, 16, 17,
\item {\em AlgGeom} (for Algebraic Geometry) for codes 11, 12, 13, 14,
\item {\em DiffGeom} (for Differential Geometry) for codes 51, 52, 53, 32, 58,
\item {\em Topology} for codes 19, 54, 55, 57, 22,
\item {\em Analysis} for codes 26, 28, 30, 41, 42, 43, 46, 47, 33, 34, 39, 40,
\item {\em PDE} (for Partial Differential Equations) for codes 31, 35, 44, 45, 49,
\item {\em DynSys} (for Dynamical Systems) for code 37 (appearing only 2000),
\item {\em Physics} (for mathematical physics) for codes 70, 74, 76, 78, 80, 81, 82, 83, 85, 86,
\item {\em Numerics} for code 65,
\item  {\em Probability} for code 60,
\item {\em Statistics} for code 62,
\item {\em Other} for codes 00, 01, 04, 97, 03, 05, 68, 73, 90, 91, 92, 93, 94.
\end{itemize}
Note that all subfields do not have the same weight, and in fact the weight of the different fields, measured in terms of total number of articles in the database, varies quite a bit over time, see Table \ref{tab:weights}. Note also that DynSys (Dynamical Systems) only appears in the period 1997-2000, since it did not have a specific 2-digit code before the 2000 revision of MSC. 

\begin{table}[h]
  \centering
  \begin{tiny}
    \begin{tabular}{|l|r|r|r|r|r|r|r|r|}
      \hline
      & 1984-88 & 1989-92 & 1993-96 & 1997-00 & 2001-04 & 2005-08 & 2009-12 & 2013-16 \\
      \hline
Other & 13.2 & 12.1 & 12.6 & 12.4 & 12.5 & 12.2 & 13.1 & 12.8 \\
Algebra & 4.6 & 4.5 & 4.7 & 4.8 & 4.6 & 6.4 & 7.4 & 7.3 \\
AlgGeom & 9.5 & 9.1 & 9.2 & 8.7 & 8.7 & 7.9 & 7.9 & 8.1 \\
DiffGeom & 12.5 & 13.4 & 12.6 & 10.3 & 8.8 & 7.8 & 8.1 & 8.5 \\
Topology & 6.0 & 5.7 & 5.0 & 4.6 & 3.9 & 3.8 & 3.5 & 3.8 \\
Analysis & 16.6 & 15.6 & 16.2 & 15.1 & 15.6 & 16.5 & 13.1 & 11.8 \\
PDE & 10.2 & 10.8 & 10.9 & 11.5 & 12.6 & 13.2 & 14.7 & 17.6 \\
DynSys & 0.0 & 0.0 & 0.0 & 2.2 & 4.5 & 4.7 & 4.1 & 4.0 \\
Physics & 3.5 & 4.6 & 4.5 & 6.3 & 6.7 & 7.5 & 6.8 & 6.8 \\
Numerics & 6.7 & 7.0 & 7.7 & 8.1 & 6.3 & 6.6 & 6.2 & 6.5 \\
Probability & 6.4 & 7.2 & 7.2 & 6.5 & 7.1 & 6.5 & 7.1 & 7.0 \\
Statistics & 10.4 & 9.4 & 8.9 & 8.9 & 8.2 & 6.5 & 7.3 & 5.2 \\
\hline
  \end{tabular}    
\end{tiny}
\caption[Weight]{Proportion of papers in different subfields}
  \label{tab:weights}
\end{table}

\subsection{The MCQ as proxy for the ``impact'' of journals}
\label{ssc:mcq}

Part of the analysis of the status of subfields of mathematics uses, or is based on, a stratification of the journals appearing in our database, in terms of their prestige or level of selectivity. The choice of the proper indicator, to be used as a proxy for the prestige of a journal, requests some care, especially in the area of mathematics.

The use of any type of bibliometric indicators has been the object of heated debates within the mathematics community. Its use for individual evaluations is often considered quite negatively. Moreover, some of the bibliometric indicators widely used, such as the Journal Impact Factor, are sometimes considered as crude and not well adapted to mathematics \cite{adler2009citation,ferrer2016impact}. In addition to general methodological objections, one main limitation of the use of IF for mathematics is that it is based on couting citations in a 2-year window which is too short for mathematics, where the cite half-life of articles tends to be much higher. This short window creates biases between subfields of mathematics depending on their citation habits. 

Perhaps for this reason, {\em Math. Reviews} also provides a measure of impact of journals, the {\em Mathematics Citation Quotient} (MCQ). It measures the mean number of citations to articles published by each journal over a 5-year period, in a selected list of journals. The longer citation window, and the selection of citation sources, make it better suited than other impact measures for mathematics. This MCQ appears much better suited to mathematics than the IF. Mathematicians tend to have a very precise idea of the ranking of journals, and anecdotical evidence (coming from e.g. evaluation panels or recruitment committees) indicates that different mathematicians usually have relatively similar assessment of the prestige of different journals, even when they specialize in different subfields. It appears -- although more research on this topic would be welcome -- that this subjective ``prestige'' of journals is much better captured by the MCQ than by the IF. 

Here we use the MCQ 2016 as an (imperfect) proxy for the ``prestige'' of journals. This view is supported by a analysis presented in Section \ref{ssc:indicators}, where it is shown that weighing each paper by a certain power of the MCQ provides an indicator that works well to identify mathematicians obtaining the highly competitive grants from the European Research Council. 

\section{First observations: Fields medals and top journals}

The first indications of the ``value'' attributed to different fields can be found by following two sources of prestige or status that are widely accepted in the mathematics community.  
\begin{itemize}
\item From the main specialties of recipients of the Fields medal, the best known and most revered prize in mathematics.
\item From the share of different fields in papers published by the ``top 5'' mathematics journals. 
\end{itemize}

\subsection{Fields medalists}
\label{ssc:fields}

A first indication of the ``status'' of subfields can be obtained by observing the specialties of the mathematicians who received the top prizes in mathematics. This is done in Table 1 of the {\em Supplementary material} for the Fields medal, by far the best known and most prestigious prize in mathematics. 

A number of observations follow from the list of subfields of Fields medalists.
\begin{itemize}
\item Algebraic and Differential Geometry, as well as Topology, tend to dominate.
\item Probability, Dynamical systems, pPartial differential equations appear in the 1990s.
\item Statistics or Numerical mathematics do not appear at all (yet). 
\end{itemize}

There are other well-known prizes in mathematics, such as the Wolf prize, \footnote{The list of recipients of the Wolf prize can be found at \url{https://en.wikipedia.org/wiki/Wolf_Prize_in_Mathematics}.} or more recently the Breakthrough Prize \footnote{See \url{https://breakthroughprize.org/Prize/3}.}, the Abel prize\footnote{See \url{https://en.wikipedia.org/wiki/Abel_Prize}.} or the Chern Medal \footnote{See \url{https://en.wikipedia.org/wiki/Chern_Medal}.}. It appears that similar conclusions would follow from considering the recipient of those prizes. Of course a different picture would emerge from {\em specialized} prizes focusing only on one or several subfields, but none of those specialized awards has yet reached the same reputation as the Fields medal.

\subsection{Papers in the ``top 3'' journals}
\label{ssc:top3}

\begin{figure}
\center{\includegraphics[width=5cm]{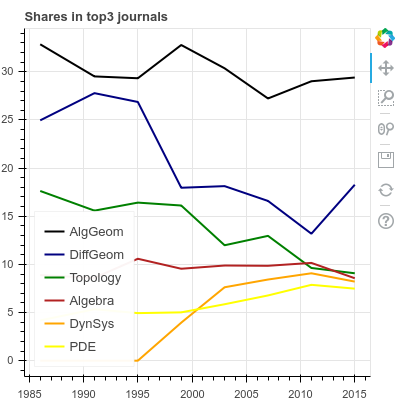}}
\caption{Share of main fields in top 3 journals}
\label{fig:top3}
\end{figure}

Other indications can be obtained from the primary and secondary MSC classifications of the papers published in the ``top 3'' journals in mathematics. There is a relatively large consensus among mathematicians on what the most selective journals are, with a standard list: {\em Annals of Mathematics}, {\em Inventiones Mathematicae}, and {\em Journal of the American Mathematical Society}. Those journals are for instance those considered as ``Top'' journals in mathematics by the ARWU ranking\footnote{See \url{http://www.shanghairanking.com/subject-survey/top-journals.html}} and therefore used for their ranking of universities in the area of mathematics. Some mathematicians would tend to add a very small number of journals to this list, in particular {\em Acta mathematica} and {\em Publications Mathématiques de l'IHES}, but those two journals are publishing a small number of articles (typically less than 10 per year each) and adding them to the list would not change the results of any bibliometric study much. 
The share of articles with primary MSC codes in the main subfields is shown in Figure \ref{fig:top3}


Those three journals are highly selective, and typically publish less than 200 articles/year {\em together} , for instance they published 139 articles in 2019. (This contrasts strongly with other scientific fields, such as Physics, where one top journal such as {\em Physical Review Letters} publishes around 4000 articles/year, while {\em Nature} published around 4000 items in 2019.) For this reason, publishing in one of the top 3 journals is seen as a significant achievements among mathematicians and can have a real impact on a mathematician's career. 

The main conclusions are  similar to those from the main areas of Fields medalists, although in a quantitatively more precise manner. Algebraic and Differential geometry dominate, and their weight is much larger than in the database of articles as a whole, as seen in Table \ref{tab:weights}. But some fields -- in particular Differential Geometry an Topology -- see a clear decrease in their share, while others -- Dynamical Systems and PDEs -- have an increasing share of the number of papers.

\section{Departments}
\label{sc:departments}

This section is focused on the disciplinary profiles of different groups of universities. We base this analysis on the well-accepted idea that departments tend to have well-established and relatively stable levels of ``status'', see \cite{clauset2015systematic} for a more refined analysis of department hierarchies in computer science, business and history.

We consider different groups of universities, and different ways of asserting their disciplinary specialization.
\begin{itemize}
\item By their scientific production  (articles published).
\item By the main fields of publication of their active mathematicians.
\item By their recruitments, that is, the disciplinary focus of mathematicians moving to those departments.
\end{itemize}
To obtain a synthetic view of the scientific focus of different universities, we defined several groups of universities, aiming towards some homogeneity within each group.
\begin{itemize}
\item Ivy-league type universities: Columbia, Cornell, Dartmouth, Harvard, Princeton, U. of Pennsylvania, Yale.
\item Technology Institutes: Caltech, GeorgiaTech, ETH, EPFL, MIT, Stanford.
\item Among the remaining universities we determined 4 groups depending on their total scientific production over the whole period: those ranked 1--10, 11--30, 31--100 and 101--300. 
\end{itemize}
Note that more refined analysis are possible, for instance differentiating US and non-US institutions, or by country, etc. We have not included such analysis here for lack of space. 

The motivation for defining those groups of universities might be clear to the reader.
\begin{itemize}
\item Ivy-league universities are generally considered to be among the top universities in the US. They also share a certain number of characteristics, for instance they are all (by definition) relatively old and well-established institutions. They are not chosen here as the ``best'' universities, but as a rather homogeneous group of universities which are presumably aiming at academic excellence. 
\item The small group of Technology Institutes considered here are also considered as ``excellent'' institutions, but with a somewhat different outlook then the Ivy-league universities. The student of those institutions tend to be mostly oriented towards engineering, and one can assume that their research activity is oriented at least in parts towards innovation and real-life applications, rather than academic research. This would lead to a specialization in subfields of mathematics somewhat different from other, more academically-oriented universities.
\item The other universities (in fact those publishing a sufficient number of papers, since we limit our study to the top 300 institutions) are broken down into four groups of decreasing total production according to an indicator that puts a relatively heavy weight on article in the top journals: each paper is weighted by its number of pages times the {\em square} of the MCQ of the journal. This weighting ensures that the institutions with a better rank tend to be those producing more papers in highly selective journals (see Section \ref{ssc:mcq}).
\end{itemize}

\subsection{Specialization of departments, by publications}
\label{ssc:dept-publis}

\begin{figure}
  \center
  {
    \includegraphics[width=4cm]{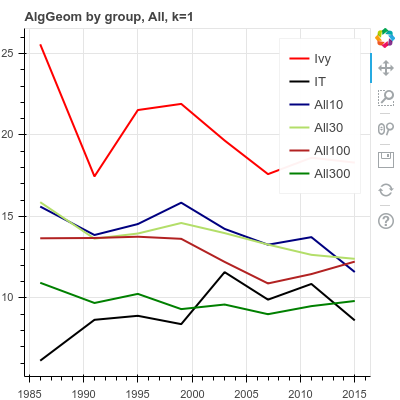}
    \includegraphics[width=4cm]{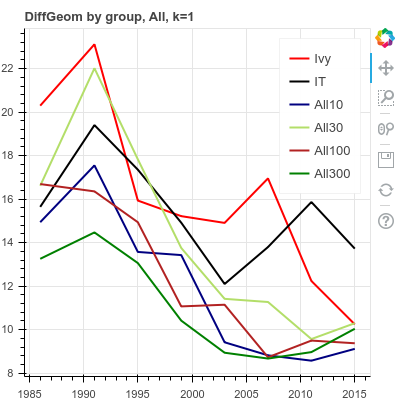}
    \includegraphics[width=4cm]{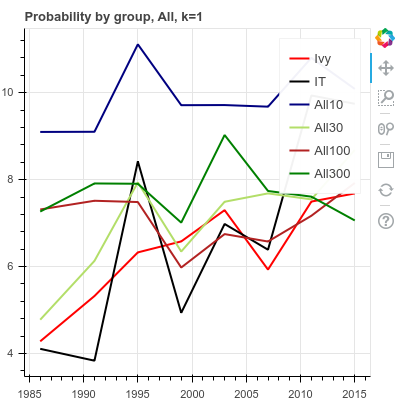}
    \includegraphics[width=4cm]{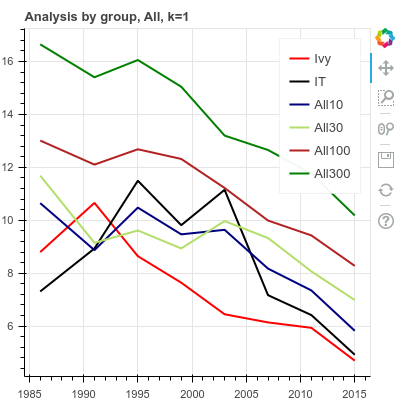}
    \includegraphics[width=4cm]{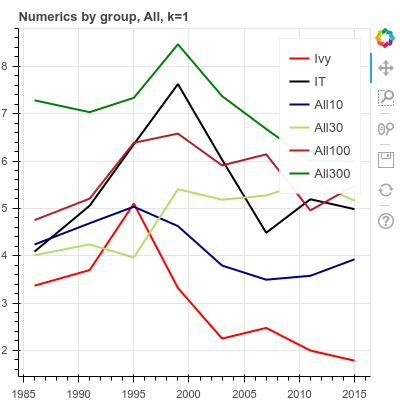}
    \includegraphics[width=4cm]{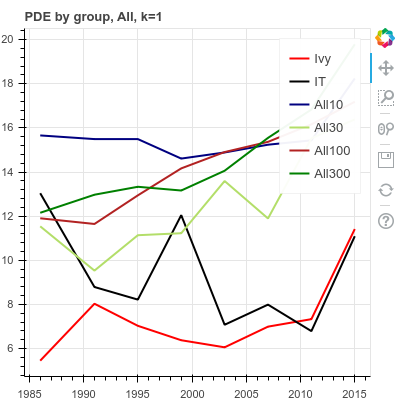}
  }
  \caption{Share of some subfields in different groups of universities}
  \label{fig:dept_shares}
\end{figure}

The weight of different subfields varies widely depending on the type of department. Some subfields have a much larger role in ``elite'' departments, while others tend to be much more present in less ambitious institutions. Figure \ref{fig:dept_shares} shows the evolution over time of the share of publication (weighted by number of pages and MCQ of the journal) in a limited group of subfields, depending on the type of department -- the full data, for all subfields, can be found in Figure 1 of the Supplementary material.


Algebraic geometry provides an prime example of a high-status field: in spite of a somewhat decreasing share, it remains dominant in Ivy league departments (approximately 17\% of papers) while it plays a much more reduced role in less prestigious departments (approximately 8\% in departments ranked 101--300). Its share is much more limited in Technology Institutes, which is to be expected since its applications to engineering are relatively limited.

Differential Geometry offers a similar ``elite'' profile at the beginning of the observation period, but the most striking phenomenon is that its share in Ivy league universities, and in fact in all groups of universities except ITs, decreases very markedly. The opposite can be observed for Probability theory, which is almost absent from Ivy league universities and ITs at the beginning of the observation period, but becomes important by the end. 

On the opposite end of the spectrum, Analysis and Numerics appear to be low-status fields  at the end of the period, with a very limited role in Ivy league departments and a much larger role in the less prestigious institutions (those ranked 101--300). Partial differential equations (PDEs) shows a similar profile, with a notable increase in status by the end of the observation period. 

It is quite apparent in the graphs that the Technological Institute have, as could be expected, a somewhat different focus than the other universities. They have less focus on Algebraic geometry and more on Numerics, as could be expected from institutions with a strong focus on engineering programs and students, but the share of Algebraic geoemtry is still larger in those institutions than in lower-ranked universities. However they do have a strong focus on Differential geometry, which cannot really be considered as an applied field.

The evolution over time of the status and share of some subfields is quite striking. For instance Analysis already appeared to be a low-status area in the 1980s, with a larger role in department with a lower ranking. By the 2010s, it had almost disappeared from the Ivy-league departments, while retaining a significant place in the departments ranked 101--300. On the opposite, Partial Differential Equations was almost absent from Ivy league institution in the 1980s, but had acquired a significant position there by the 2010s. 


\subsection{Specialization of departments, by active mathematicians}
\label{ssc:special}

Another way to look at the specialization of departments is through the specialties of their researchers. 

One obstacle in this direction is that articles appearing in our database are not necessarily written by members of mathematics departments. This is particularly true in some applied fields like statistics or numerical modelling, where authors might not be affiliated to a department of Mathematics (or Mathematics and statistics, or even of Statistics). 

We intend here to measure the number of mathematicians, in the sense of scientists who devote a significant portion of their activity to mathematics research over a long period of time, and therefore prefer excluding PhD students who leave academia after their thesis, or scientists of other fields producing only occasionally a paper that is related to mathematics and included in the data we use. For this reason, we only consider authors who have over their lifetime produced a total of 100 pages in the journals we consider, each page being weighted by the MCQ of the journal in which it is published and divided by the number of authors (reflecting the relative contribution of the author).

For each period, each mathematician is attributed a primary subfield, determined as the most common field of their paper (weighted by MCQ and number of pages). Note that the primary field of mathematicians can change from one period to another. 

\begin{figure}
  \center{
    \includegraphics[width=4cm]{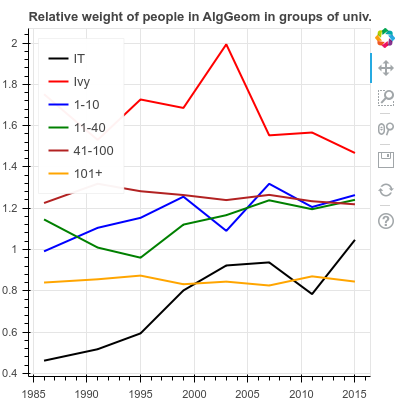}
    \includegraphics[width=4cm]{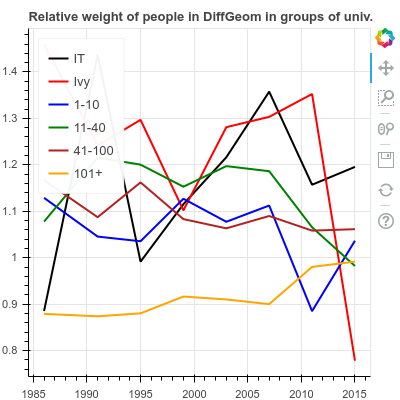}
    \includegraphics[width=4cm]{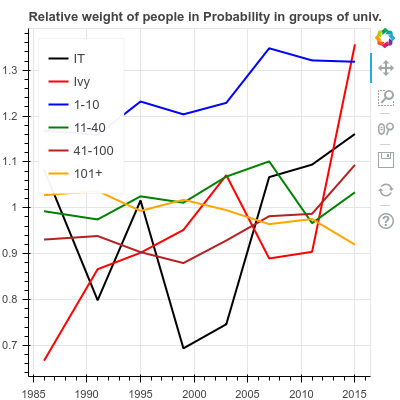}
    \includegraphics[width=4cm]{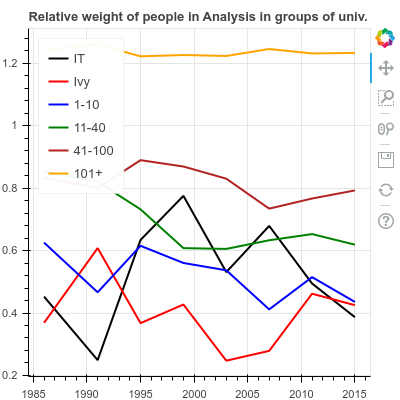}
    \includegraphics[width=4cm]{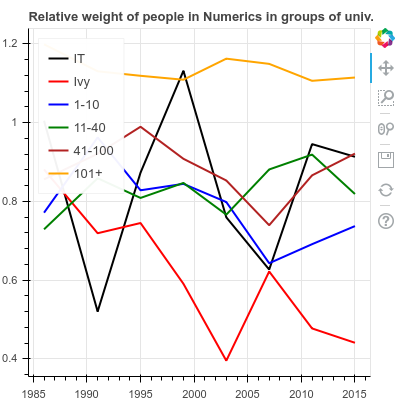}
    \includegraphics[width=4cm]{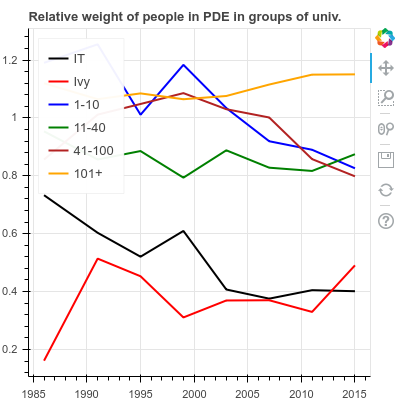}
  }
  \caption{Relative weights of mathematicians in selected fields depending on groups of universities}
  \label{fig:relative}
\end{figure}

Figure \ref{fig:relative} shows the relative weights of a selection of subfields in different types of universites, over time. Here a relative weight of 1.2 for Algebra in ITs, for instance, would mean that there are 120 mathematicians with main focus on Algebra in ITs when the expected number (knowing the total number of mathematicians in ITs and the proportion of those specializing in Algebra overall) is 100.


Figure \ref{fig:relative} shows the resulting data for the same subfields as in Figure \ref{fig:dept_shares}. Again, the full data for all subfields is available in the Supplementary material. 

The results are somewhat different from that presented in Section \ref{ssc:dept-publis} since different subfields tend to publish more papers than others, on average (see below). Comparing numbers of active mathematicians gives a better view of the focus put on different subfields by departments. The results are also quite clear, for instance Algebraic geometry appears again as a high-status fields, with a strong over-representation in Ivy league departments and under-representation in departments ranked 101-300. Differential geometry is also a ``high-status'' field, but with a clear decrease in its position at the end of the period, while Probability went from low status to high status over the period. Analysis, Numerics and PDEs appear as low-status fields, with a strong under-representation in top departments and a strong over-representation in 100+ departments.


\subsection{Recruitments}
\label{ssc:recruitments}

The graphs shown in the previous two sections show that the weight of some subfields in some departments --- for instance Differential geometry, or Probability theory --- varies very quickly. In this section and the next we analyze the mechanism for those rapid variations, first by considering in what subfields recruitments are made, and in the next section to what extend mathematicians change from one subfield to the other. One key results is that both departments and individual mathematicians tend to be quite dynamic, but that the departments with a higher ``status'' (and their members) move more quickly. 

\medskip

\begin{figure}
  \center{
    \includegraphics[width=3.8cm]{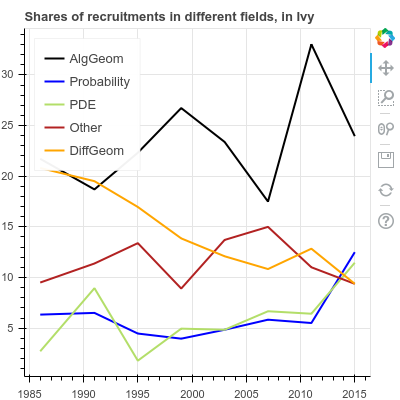}
    \includegraphics[width=3.8cm]{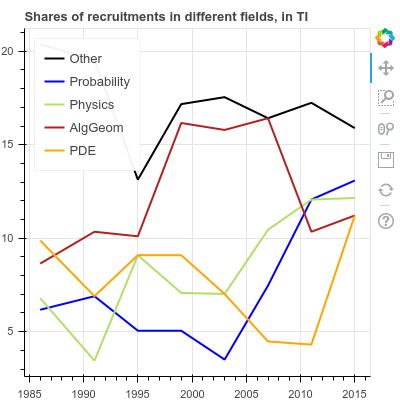}
    \includegraphics[width=3.8cm]{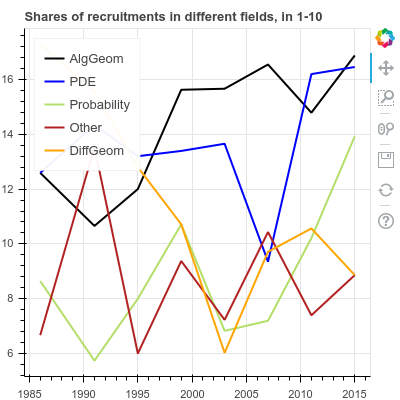}\\
    \includegraphics[width=3.8cm]{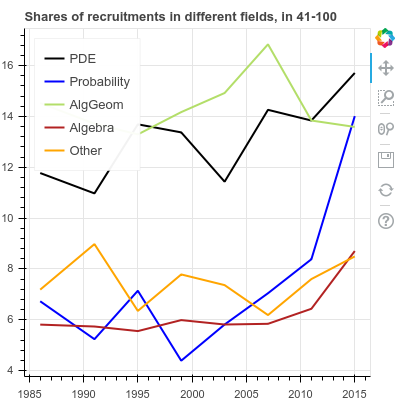}
    \includegraphics[width=3.8cm]{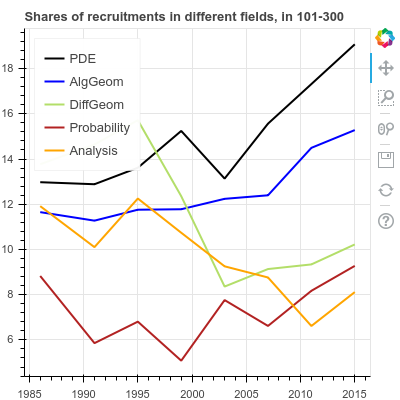}
    \includegraphics[width=3.8cm]{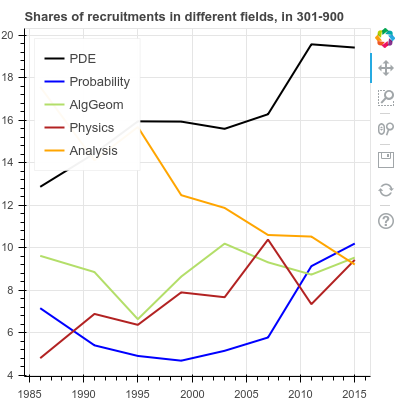}
    }
    \caption{Share of recruitments in different fields for different groups of universities}
    \label{fig:share_hires}
\end{figure}

To extend the analysis and obtain a more dynamics view, one can look at the proportion of different fields among mathematicians recruited in different groups of universities, see Figure \ref{fig:share_hires}.


Figure \ref{fig:share_hires} shows the proportion or recruitments (defined as arrivals of new authors in an institution, at least 2 years after the PhD). It displays interesting differences between the main fields where different groups of institutions choose their new recruits. Top departments (Ivy league and top 10 ``other'' departments) continue to give a strong preference to Algebraic Geometry, and a decreasing place is given to Differential Geometry. Partial Differential Equations and Probability see their role increasing in basically all groups of universities, while Analysis and Algebra remain important in lower-ranked institution (ranked 301 and below).


\begin{figure}
  \center{
    \includegraphics[width=3.8cm]{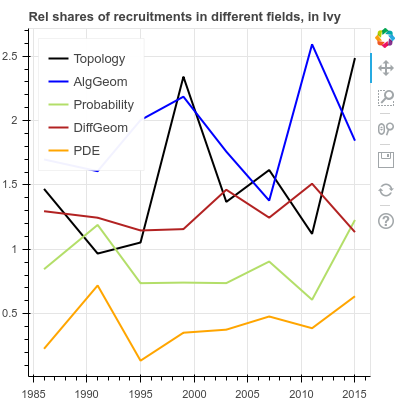}
    \includegraphics[width=3.8cm]{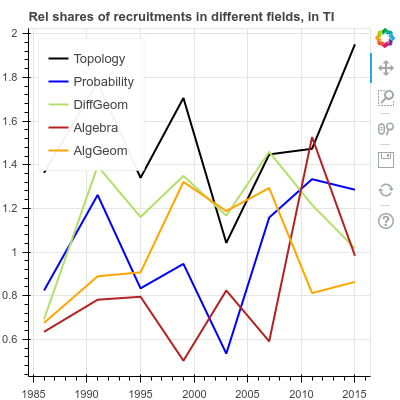}
    \includegraphics[width=3.8cm]{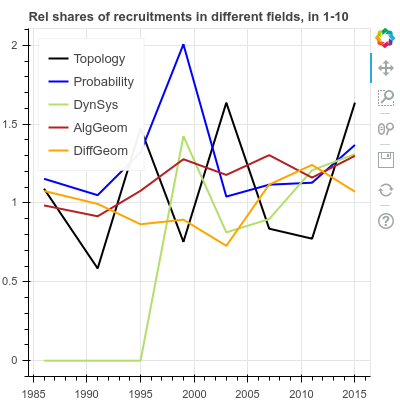} \\
    \includegraphics[width=3.8cm]{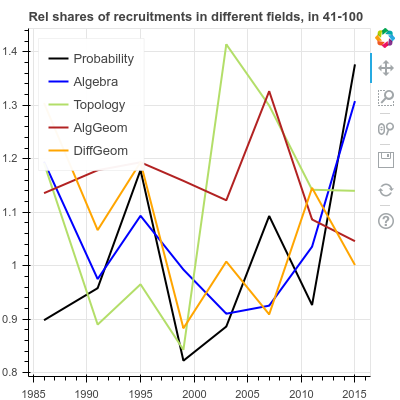}
    \includegraphics[width=3.8cm]{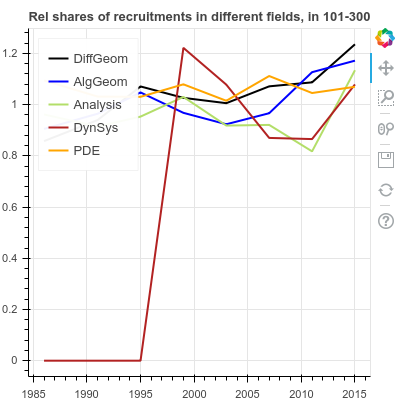}
    \includegraphics[width=3.8cm]{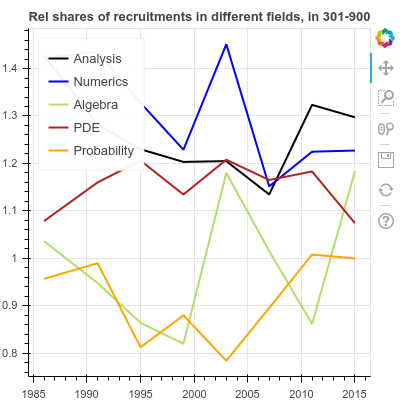}
    }
  \caption{{\em Relative} share of recruitments in different fields for different groups of universities}
    \label{fig:share_hires_rel}
\end{figure}

Figure \ref{fig:share_hires_rel} is somewhat simpler to interpret as it shows the {\em relative} shares of different fields in the recruitments of different groups of institutions\footnote{Note that the subfields Statistics, Physics and Others were not considered here, again because they appear to reflect in an important manner recruitments made in non-mathematics departments.}. For each group, only the 5 subfields with the largest relative share of recruitments at the end of the period appear. There is again a very clear distinction between top institutions (Ivy league, ITs, institutions ranked 1--10 in our list) where ``high-status'' fields such as Topology and Algebraic Geometry dominate, with a relative weight close or larger than 2, together with ``new'' highly regarded fields such as Probability, while Differential Geometry keeps a significant but slightly decreasing place. On the opposite, in less regarded institutions (ranked below 301 and even more below 901) the recruitments disproportionately favor Numerics, Analysis or and PDEs. 

\subsection{Individual mobility between subfields}
\label{ssc:mobility}

It can be noted that the recruitments made in different fields are not sufficient to explain the important variation in the weight of different fields seen for instance in Figure \ref{fig:dept_shares}. Another possible explanation can be found in differing {\em attrition rates}: at all stages of their careers, a certain proportion of mathematicians tend to stop publishing, see \cite{drs}, and this rate could vary from one field to another.

\begin{figure}
  \center{
    \includegraphics[width=16cm]{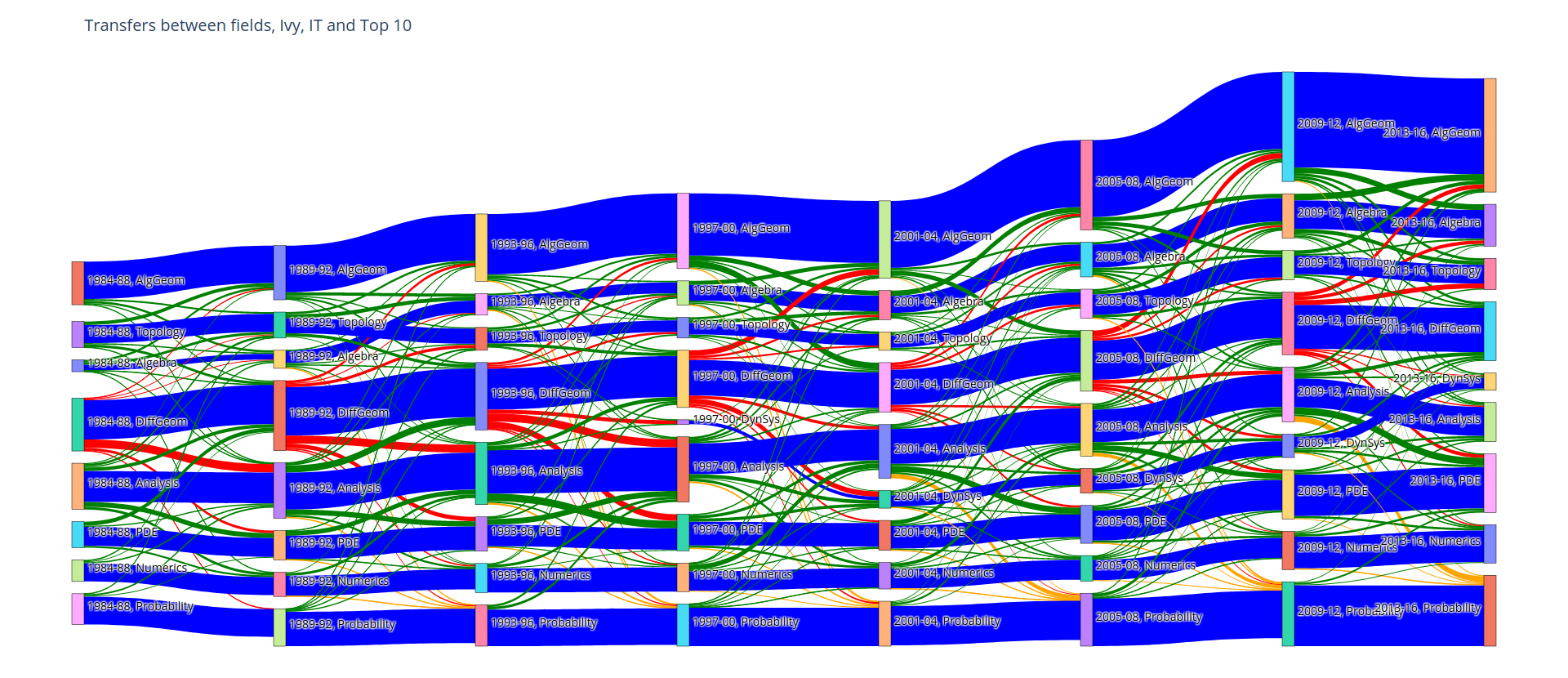}
    \includegraphics[width=16cm]{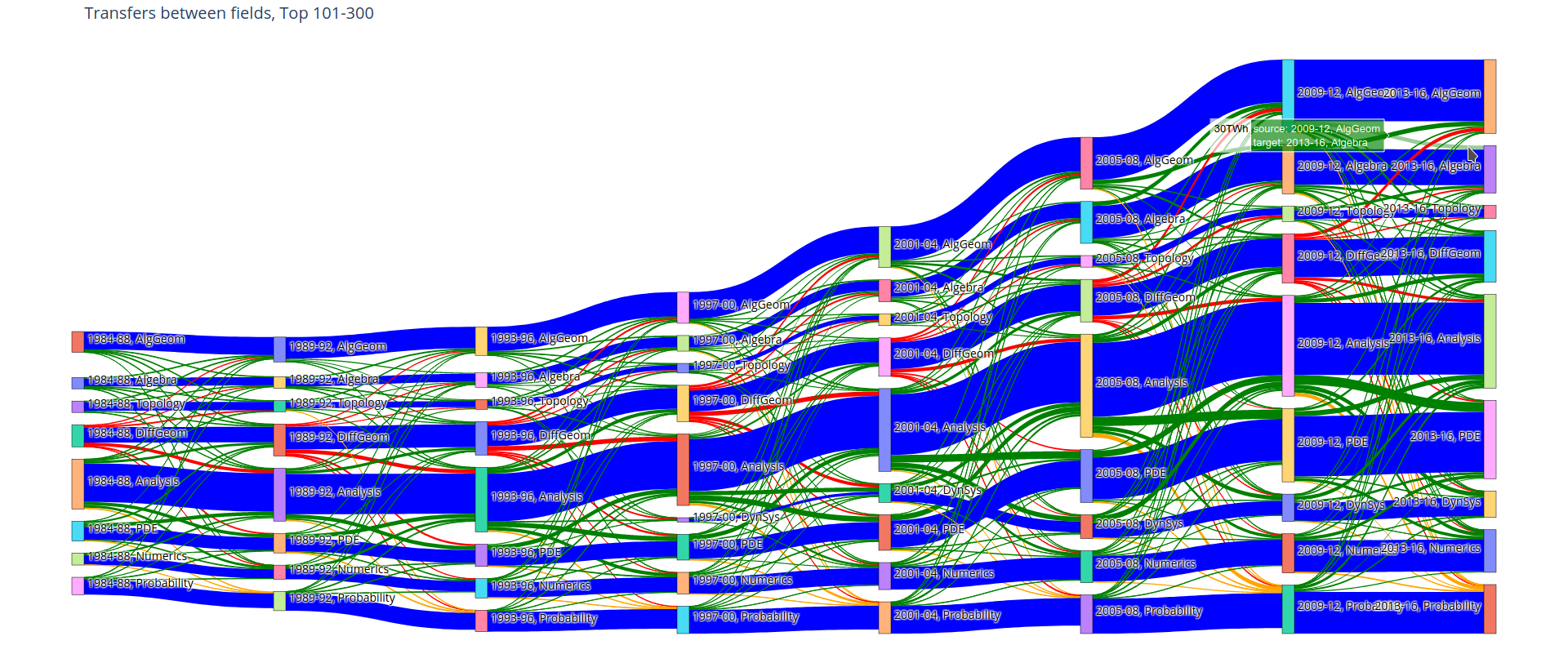}
    }
  \caption{Transfers between subfields. Top: Ivy, IT and top 10. Bottom: 101--300.}
    \label{fig:sankey}
\end{figure}

Another important element, however, is the mobility of mathematicians between fields. Figure \ref{fig:sankey} shows the mobility in and out of some ``core'' fields of mathematics (excluding Statistics, Physics and Other, where an important part of publications is not from members of mathematics departments). More complete data can be found in tables in the Supplementary material. Here we record only the number of mathematicians who changed their main field of publication from one period to the next (without considering incomers who did not publish before, or outgoers who stop publishing), where the ``main field'' is determined as the most common field defined by primary classification of papers published in a given period. Some remarks that can be made are:
\begin{itemize}
\item The in and out mobility is quite large for most fields and most periods -- this can be explained by the fact that a significant proportion of mathematicians are at the interface between two fields, and can lean towards one side or the other from one period to the next.
\item The level of mobility tends however to decrease with decreasing rank of universities -- mathematicians at ``better'' universities tend to change their topics more than those at ``lower'' institutions.
\item There are very significant imbalances between incoming and outgoing flow in some periods for some fields, for instance out of Differential Geometry in the periods 1997--2000 and 2001--2004 (out flow in red) and towards Probability in 2009--2012 (in flow in orange).\end{itemize}

\medskip

Figure \ref{fig:sankey} hints at the role of individual strategies of researchers, who tend to leave some subfields of decreasing status and move towards those of increasing prestige. It is conceivable that publishing in the most selective journals is perceived as easier in some subfields than in others. However, moving to a new field can be very costly in terms of time and energy, especially when moving to subfields where an important background is needed, which tends to be the case of most high-status subfields (such as Algebraic geometry or Topology). 

However the transfers of authors from one subfield to another, as seen in Figure \ref{fig:sankey}, might not be sufficient to explain the variations in the number of publications in some subfields in some groups of universities. The priority given to some subfields in recruitments can also play a role, see Section \ref{ssc:recruitments}. Other explanations can be found in the attrition rate (the proportion of mathematicians who stop publishing), which also appears to differ markedly between subfields and over time. We do not include an analysis of this phenomenon here since it would need to be relatively complex and various factors need to be taken into consideration (e.g. the different numbers of PhD students in different subfields, relations to industry, risk of ``missing'' some publications in some fields more than others, etc). 

\section{Journals}
\label{sc:journals}

\subsection{Discipinary focus of journals}
\label{ssc:disciplinary}

As mentioned in the introduction, another way to look at the ``status'' of subfields of mathematics is to consider whether top-ranked journals tend to publish more often papers in some subfields than in others.

To perform this analysis, we use the MCQ provided by {\em Math Reviews} as a proxy for journal ``quality'' (specifically, we use the 2016 value of the MCQ). One should note however that this analysis is somewhat complicated by the difference in publishing and in citing behaviors between subfields of mathematics. Typically, papers in some subfields (for instance Partial differential equations or Statistics) tend to be cited faster, and therefore more often in the 5-year timeframe of the MCQ, than in others (for instance Algebraic Geometry). As a consequence, journals which are specialized in some fields or just tend to publish more papers in papers in certain subfields can have a higher MCQ than would be estimated from their ``quality'' as perceived by mathematicians.

\begin{figure}
  \center{
    \includegraphics[width=3.8cm]{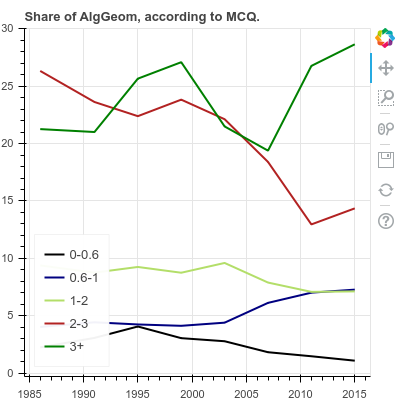}
    \includegraphics[width=3.8cm]{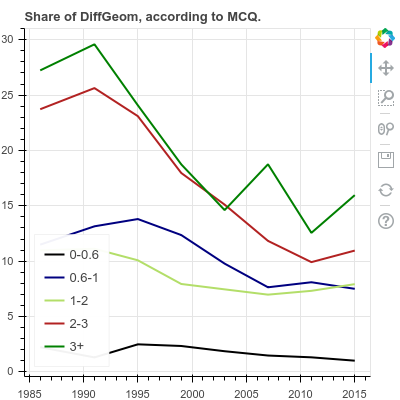}
    \includegraphics[width=3.8cm]{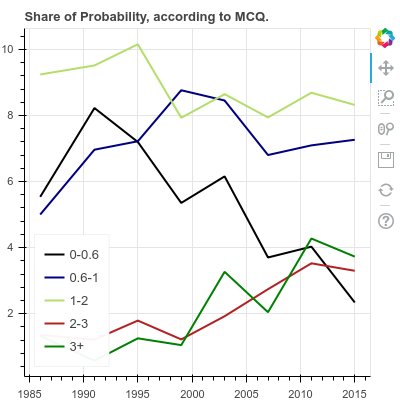} \\
    \includegraphics[width=3.8cm]{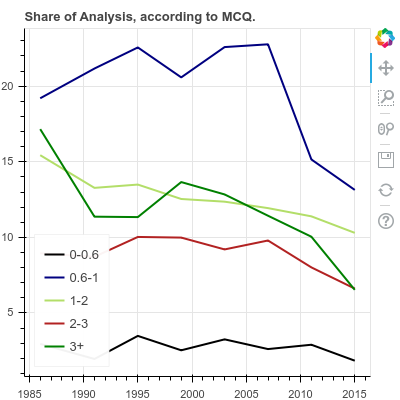}
    \includegraphics[width=3.8cm]{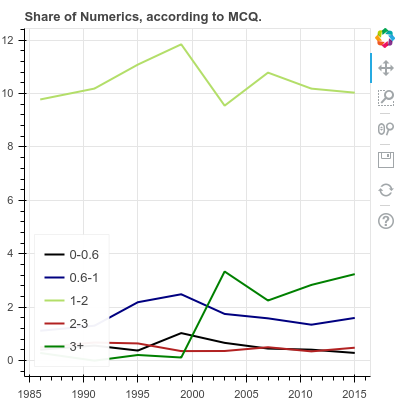}
    \includegraphics[width=3.8cm]{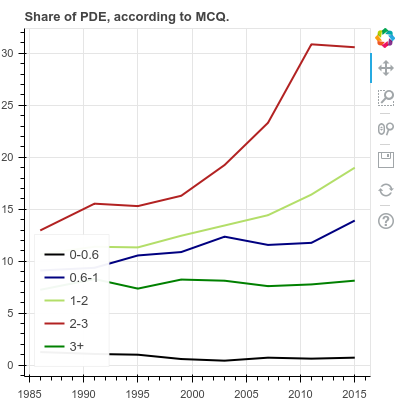}
  }
  \caption{Share of some fields according to journal MCQ}
  \label{fig:share_journals}
\end{figure}

\begin{figure}
  \center{
    \includegraphics[width=3.8cm]{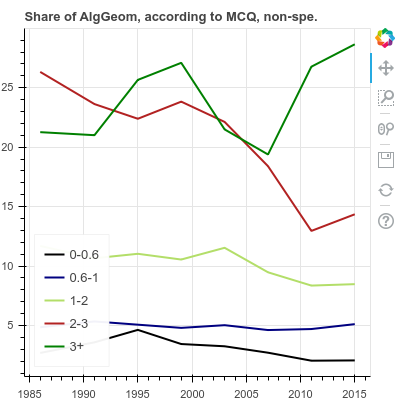}
    \includegraphics[width=3.8cm]{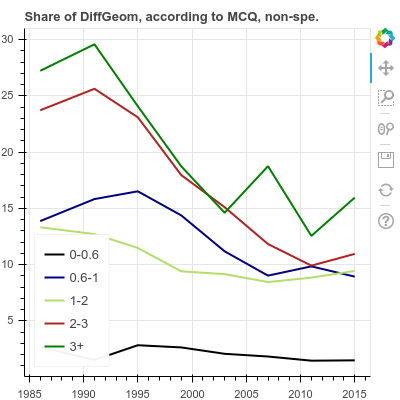}
    \includegraphics[width=3.8cm]{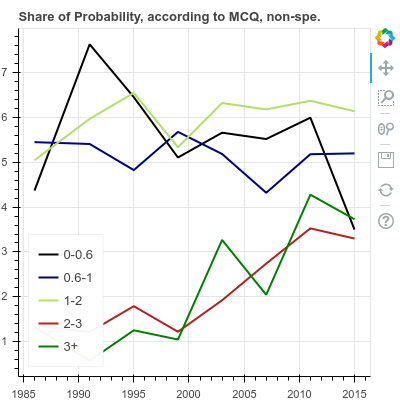} \\
    \includegraphics[width=3.8cm]{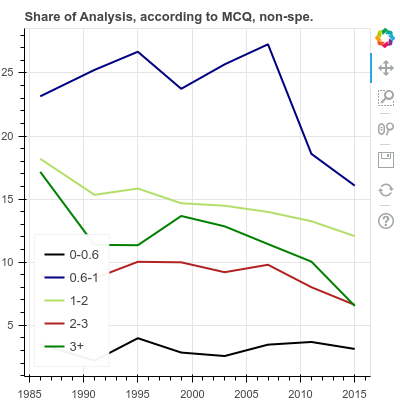}
    \includegraphics[width=3.8cm]{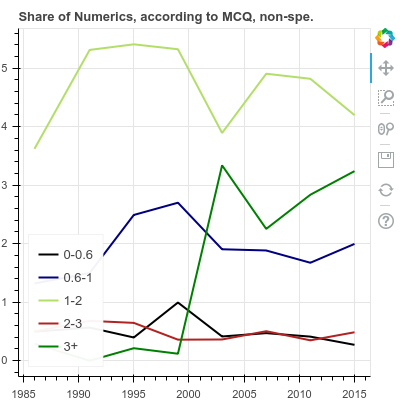}
    \includegraphics[width=3.8cm]{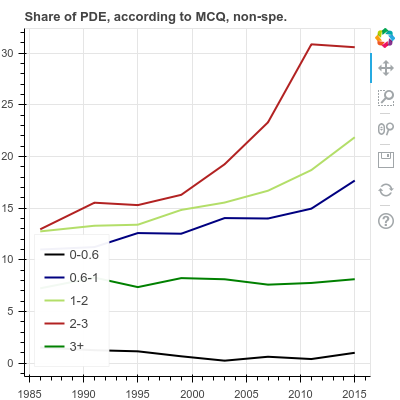}
  }
  \caption{Share of some fields according to journal MCQ, excluding specialized journals}
  \label{fig:share_journals_spe}
\end{figure}

To avoid this bias, we provide two figures: Figure \ref{fig:share_journals} shows the weight of different subfields in groups of journals depending on their MCQ, while Figure \ref{fig:share_journals_spe} provide the same data but excluding journals with a high degree of specialization. (The degree of specialization used here is computed as the sum of the squares of the relative weight of each subfield among papers published by the journal, where the relative weight in a given field is 1 if the proportion of papers in that field is the same as in the whole database). Although the two figures are quite comparable, some specific artifacts tend to disappear in Figure \ref{fig:share_journals_spe}.

Those figures quite precisely support the idea of a ``ranking'' of subfields as already seen in Section \ref{ssc:top3} and in Section \ref{sc:departments}, as well as the {\em dynamics} of this ranking. Algebraic Geometry is heavily over-represented in journals with a high MCQ, while Differential Geometry is over-represented, but with a declining share in top journals. On the other hand, Analysis for instance is under-represented in top journals, with a declining share, while the ``status'' of Probability is clearly increasing.

The figures presented here only concern the same six subfields as above, more complete data is available in the Supplementary materials.


The general picture is again that Algebraic Geometry appears as a high-status field, with a much higher weight in high-MCQ journals, while Differential Geometry has a high but decreasing status -- its weight in high-MCQ journal decreases markedly, while its weight in lower-impact journals decreases more slowly -- and the status of Probability theory increases. On the lower side of the pictures we see that Analysis, Numerics and PDEs tend to be of lower status, with a clear increase for PDEs. Removing highly specialized journals changes the picture to some (limited) extend, for instance it appears that the relatively high weight of Numerics in journals with MCQ between 1 and 2 is largely due to specialized journals. 

\subsection{A finer look at the evolving weights of subfields}

The ``subfields'' introduced above can be analysed further by considering separately the main 2-digit MSC classification codes associated to each article by {\em Mathematical Reviews}. This more detailed analysis is interesting for instance for the field {\em DiffGeom} considered above, which is defined by merging 5 different 2-digit MSC codes: 32 (Several Complex Variables And Analytic Spaces), 51 (Geometry), 52 (Convex And Discrete Geometry), 53 (Differential Geometry) and 58 (Global Analysis, Analysis On Manifolds).
 
\begin{figure}
  \center{
    \includegraphics[width=4cm]{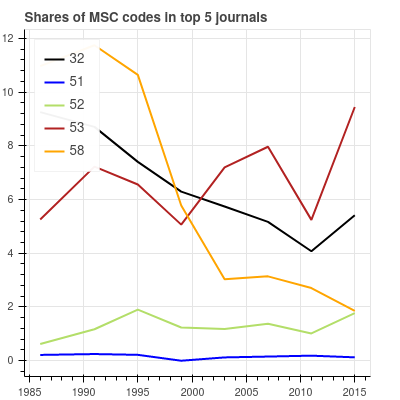}
    \includegraphics[width=4cm]{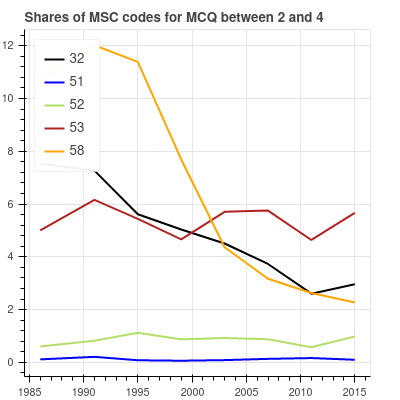}
  }
  \caption{Share of 2-digit MSC codes in {\em Differential Geometry} in top 5 journals and in journals with MCQ between 2 and 4}
  \label{fig:finer}
\end{figure}

Figure \ref{fig:finer} shows a finer analysis of the share over time of those 2-digit MSC codes in high-impact journals. It shows quite cleary that the decline of Differential Geometry is due entirely to two 2-digit codes: 32 (Several Complex Variables And Analytic Spaces) and 58 (Global Analysis, Analysis On Manifolds). A similar detailed analysis could be performed for other subfields, to get a better view and understanding of the evolving share of different subfields.

\section{Publishing habits and status}
\label{sc:habits}

We have considered so far data that can reasonably be related to the ``status'' of a subfield, such as its relative weight in different departments or in different journals. In this section, we consider differences in publishing habits between different fields, first in terms of lengths of papers, second in terms of number of co-authors. It is quite striking that in both cases there appears to be a strong relation between the ``status'' of a subfield as seen above and the publishing habits of its authors. The results presented here can be compared to, and contrasted with, those in \cite{braxton1996variation,fanelli2013} concerning comparisons between different fields of science.

In the next section we will see how those differences in publishing habits can hint at explanations of the differences in ``status'' between subfields.

\subsection{Length of papers}

\begin{figure}
  \center{
    \includegraphics[width=4cm]{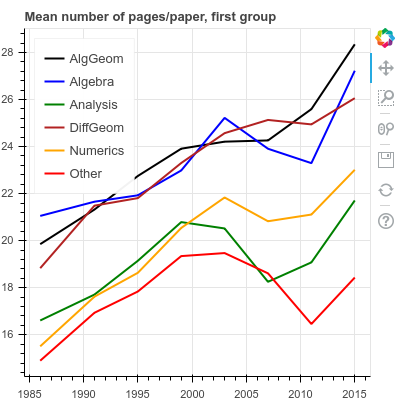}
    \includegraphics[width=4cm]{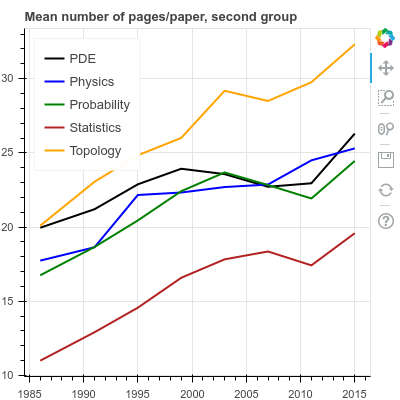}
  }
  \caption{Mean number of pages/article depending on the field}
  \label{fig:pages}
\end{figure}

The first parameter that we can consider is simply the average length of papers, see Figure \ref{fig:pages}.
Two remarks should be quite clear from the graphs.
\begin{enumerate}
\item There is a general increase in the average length of papers, as already documented for mathematics in \cite{drs}.
\item There is a remarkably direct relation between the ``status'' of subfields as seen in Section \ref{sc:departments} and in Section \ref{sc:journals}, on one hand, and the average lengths of papers. Papers tend to be significantly longer in ``high-status'' fields like Algebraic geometry, Differential Geometry -- where the length of papers compared to Algebraic Geometry decreases over time -- or Topology, and shorter in ``low-status'' fields such as Numerics or Analysis.
\end{enumerate}

It should be noted that the second point is in strong contrast  with the {\em inverse} relation between consensus level of a field and the average length of papers, observed when comparing different fields of science, see e.g. \cite{fanelli2013}. This difference could be explained by the nature of papers in mathematics, which might play a different role than in other fields of science. While in many fields a paper reports on the results of research, for instance of a series of experiments, a paper in mathematics is expected to contain a full proof of its main results -- the paper {\em is} the research. Longer papers could therefore correlate with a higher level of complexity in the proofs. 


\subsection{Number of co-authors}

\begin{figure}
  \center{
    \includegraphics[width=4cm]{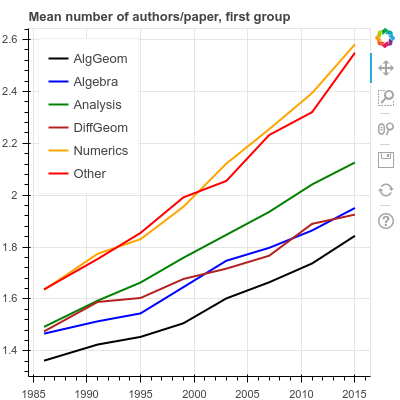}
    \includegraphics[width=4cm]{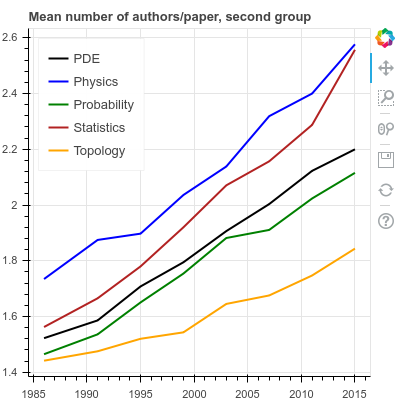}
  }
  \caption{Mean number of authors/article depending on the field}
  \label{fig:coaut}
\end{figure}

The second parameter is the mean number of co-authors of papers, as see in Figure \ref{fig:coaut}. Here too, there is a very significant rise in the mean number of co-authors over time (as also documented for mathematics in \cite{drs}. 
Moreover, the higher the ``status'' of a field, the less co-authors a papers has on average: at the end of the period of study, papers had on average less than 2 co-authors in Algebraic Geometry, Algebra, Differential Geometry and Topology, but more than 2,5 in Numerics, Others, Physics and Statistics.

Note that the interpretation of those data might require some care. The number of co-authors tends to be higher in more applied areas, and one possible source of difference might be differences in habits towards co-authorship, for instance it is more common in some subfield than in others for a PhD advisor to co-author papers with her/his PhD students.


\section{Scientific productivity in different fields}

It should come as no surprise, given the varying weight of different subfields as seen in Section \ref{sc:departments} and in Section \ref{sc:journals}, that the ``productivity'' of mathematicians in different fields vary, with the amount of variation depending on the indicator of production which is considered. 

\subsection{What are proper indicators of ``production''?}
\label{ssc:indicators}

Before comparing scientific production between subfields of mathematics, it is necessary to define a proper indicator of ``quality'' of scientific production. Clearly many indicators can be imagined: one could for instance count the number of pages published, or weight this number in many different ways. One therefore needs a measure of how ``reasonable'' an indicator is.

For the purpose of this study, the most relevant indicators are those which best fit with the assessments of experts in the field of mathematics. To assess the quality of possible indicators, we use a large set of projects positively evaluated by experts: the projects funded by the {\em European Research Council} (ERC) in the area of mathematics (panel PE1). This is a total of 389 projects, in three categories (``Starting'', ``Consolidator'' and ``Advanced'' grants) corresponding to different age groups, always with only one principal investigator. Since the main data on the funded projects is freely available (including the identity of the PI), it is possible to estimate to what extend a given bibliometric indicator correlates well with the obtention of ERC grants. 

\begin{table}[h]
  \centering
  \begin{tabular}{|l|r|r|r|r|r|}
    \hline
Group & MCQ$^0$ & MCQ$^1$ & MCQ$^2$ & MCQ$^{2.5}$ & MCQ$^3$  \\
    \hline
top 50 & 5 & 7 & 11 & 10 & 10 \\
top 100 & 13 & 16 & 16 & 16 & 15 \\
top 200 & 29 & 34 & 33 & 36 & 32 \\
top 500 & 54 & 62 & 73 & 69 & 63 \\
top 1000 & 93 & 109 & 110 & 113 & 113 \\
top 5000 & 241 & 257 & 267 & 266 & 261 \\
    \hline
  \end{tabular}
  \caption{Number of ERC grantees among top producers according to different indicators}
  \label{tab:mcq}
\end{table}

We considered what type of bibliometric indicator to use in this light. Specifically, we tried to estimate to what extend different indicators based on the number of pages, co-authors and MCQ of the journal. We particularly focused on the best way to take into account the MCQ of journals. For different indicators, we computed how many ERC grantees would be within the top 50, 100, 200, 500, 1000 and 5000 authors in our database (over all years). The results are shown in Table \ref{tab:mcq} using as indicator the sum, for each author, over all the papers in our database of the number of pages times a power of the MCQ 2016 of the journal where it was published: from power $0$ (first column) to power $3$ (rightmost column). As can be seen, the best results are obtained, depending on the line, by a power $2$ or $2.5$ of the MCQ. In addition:
\begin{itemize}
\item Taking into account the number of pages leads to clearly better results than just counting the number of papers.
\item Taking into account the number of co-authors (for instance by dividing the weight attributed of a paper by the number of co-authors) leads to worse results.
\end{itemize}

In the following section we use one of the indicators that gives the best fit -- each paper is weighted by the product of the number of pages by the square of the MCQ of the journal -- to give an estimate of the ``production'' of mathematicians in different fields. 

\subsection{Differing production between fields}
\label{ssc:differing}

\begin{figure}
  \center{
    \includegraphics[width=3.8cm]{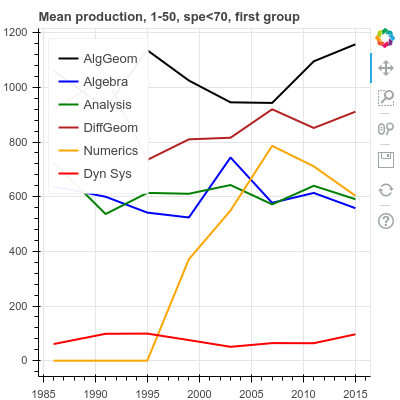}
    \includegraphics[width=3.8cm]{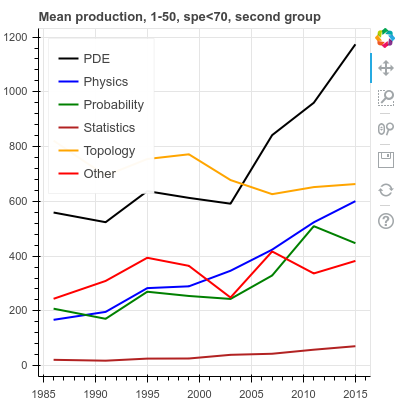}
    \includegraphics[width=3.8cm]{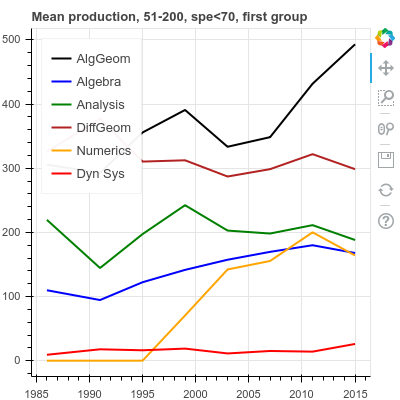}
    \includegraphics[width=3.8cm]{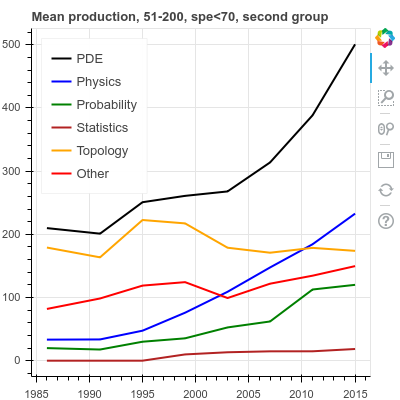}
    \includegraphics[width=3.8cm]{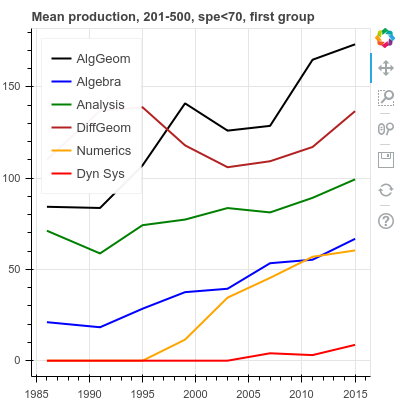}
    \includegraphics[width=3.8cm]{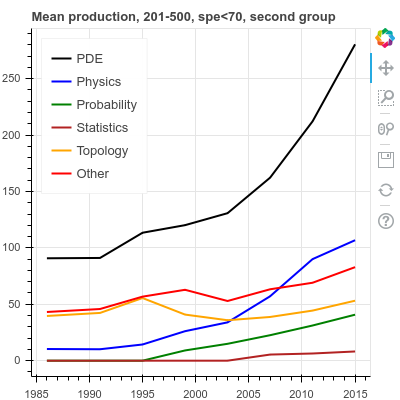}
    \includegraphics[width=3.8cm]{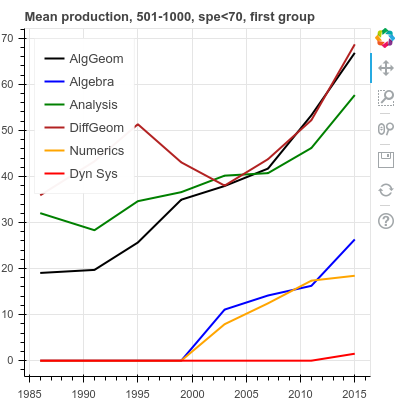}
    \includegraphics[width=3.8cm]{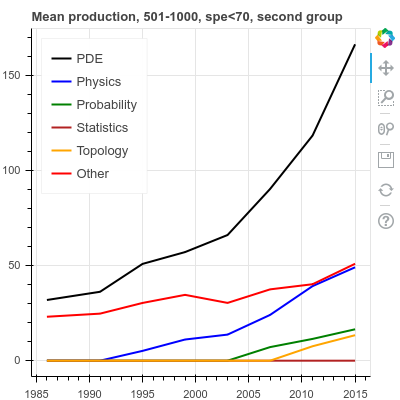}
  }
  \caption{Mean production (pages times MCQ$^2$ shared between authors) of different groups of authors depending on the field}
  \label{fig:meanProd}
\end{figure}

Using the indicator above, it is possible to measure the ``productivity'' of mathematicians in different subfields. The results are presented in Figure \ref{fig:meanProd}. For each subfield, the graphs present the average ``production'' of the mathematicians in a group defined through the ``ranking'' of all authors by production in a given period, for instance the first two graphs are for the top 50 authors in each field, in each period, the graphs number 3 and 4 for the authors ranked between rank 51 and 200, etc. In each case we split the subfields in two groups for clarity.

To avoid biases related to specialized journals having a relatively high MCQ due to faster citation rates in some fields, the data here was limited to journals having a specialization index (as defined in Section \ref{ssc:disciplinary}) at most 70. This excludes a relatively small number of journals with a higher level of specialization.

The data is quite reminiscent to the results presented above concerning the relative weights of different fields in top 5 journals or with journals with higher MCQ (as could be expected) but also with recruitments in higher status departments. The mean production is highest in fields that were identified as ``high-status'', in particular Algebraic and Differential geometry, lower in statistics.

Two relatively striking phenomena however can be noted in the {\em second part} of the time frame considered here.
\begin{itemize}
\item Authors in Partial differential equations have a high production -- as high as those in ``high-status'' fields for highly productive authors (top 50 of each field) and much larger among less productive authors (more than twice higher, in the last period, for authors ranked 501--2000).
\item Some authors in Numerics also have a relatively high ``productivity'', as high in the last period, in the group of most productive authors, as for Algebra or Analysis. 
\end{itemize}
Those phenomena should be considered with some care, since they could to some extend be explained by differences in the publication and citation practices in different fields.

\section{Analysis}
\label{sc:explanations}

\subsection{How can the ``status'' of subfields be explained?}
\label{ssc:explanations}

The data presented above leads to a simple question: what can explain the preference of top departments/journals for certain fields?
We can propose some possible answers.
\begin{itemize}
\item {\em Some fields are more useful or important for applications than others.}
  This does not seem to be the correct explanation since top departments tend to have
  preferences which are {\em opposite} to this orientation. Funding agencies 
  tend to prefer those application-oriented fields, which should provide departments
  another strong reason to develop them, but that's not what's happening.\footnote{As an example, in 2011, EPSERC decided to fund PhD scholarships only in the areas of statistics and applied probability, see \url{http://blogs.nature.com/news/2011/09/uk_mathematicians_protest_fell.html}.}
\item {\em Maximizing external funding.} For the same reason, this explanation does not seem to resist scrutinity, since funding agency tend to prefer more applied fields, which also tend to have lower ``status'' as determined above.
\item {\em The ``impact'' of a field}, as measured by standard indicators such as the number of citations that a paper can be expected to attract. Here again the data seems to clearly invalidate this explanation, since the ``high-status'' fields, such as Algebraic Geometry, are typically those where the impact of papers (at least in the 2 years after publication considered by the Impact factor) is the {\em lowest}, while fields with much higher citation impact, such as Statistics of Numerics, appear as having a much lower ``status''.\footnote{A simple experiment confirming this view can be performed by checking the number of citations of say the 10th most cited author in {\em google scholar} with a given ``label''. On June 21, 2020, this yields for instance 10 156 citations for the 10th most cited author with the label ``algebraic\_geometry'', vs 166 049 citations for the 10th most cited author with the label ``statistics''.}
\item Departments tend to maximize another type of impact, closer to the indicator used in the previous Section \ref{ssc:indicators}, and to the assessment that other mathematicians have of the relevance and importance of results. This explanation seems broadly validated by comparing the results of Sections \ref{ssc:special} and \ref{ssc:recruitments} and Section \ref{ssc:differing}, but it is somewhat circular and leads to another very close question: why do the results in some field appear more relevant or important to mathematicians than in others? 
\item {\em Consensus level.} One can be tempted to extend to mathematics a main explanation of the ``hierarchy of science'', namely, by a difference in {\em level of consensus} between different subfields. This explanation however runs into an obvious difficulty: the consensus level tends to be extremely and uniformly high across mathematics, since all published papers are expected to contain full and complete proofs and therefore to only present results that are unquestionably true. 
\item {\em Focus.} This is the main explanation which we would like to put forward. Preferred fields tend to correspond to those where there is a strong and shared focus on well-identified problems. This can be reflected in some bibliometric data:
  \begin{itemize}
  \item  Longer papers, because more efforts and technical developments are often necessary to make progress on a problem which is well identified and on which other experts already tried to make progress.
  \item Less co-authors/paper, for a similar reason -- depth and technical difficulties tend to increase the cost/benefit ratio of collaboration (too much time is spent explaining new ideas to collaborators). 
  \end{itemize}
  We present in the next section some (limited) data supporting this assumption, showing that subfields where the word {\em conjecture} is used more often tend to be higher-status. 
\item {\em Relation to teaching.} Another possible hypothesis is that some subfields are more relevant to teaching, in the sense that their researchers tend to be more involved in or dedicated to the teaching activities of their departments. We don't have any data here to support or contradict this hypothesis.
  \item {\em Some fields offer better job perspectives.} This is another simple explanation that does not appear to resist scrutinity. In fact, a quick search by keywords on the available positions on {\em mathjobs.org}, which is perhaps the main source of job advertisements in mathematics (whether academic or non-academic) seems to indicate that {\em more} jobs are available in lower-status subfields such as Statistics or Numerical analysis than in ``high-status'' fields such as Algebraic geometry.
\end{itemize}

In the rest of this section we discuss evidence supporting the idea that the level of {\em focus} of a subfield of mathematics is related to its position in the ``hierarchy of subfields''.

\subsection{The {\em focus} of a field}
\label{ssc:focus}


The notion of focus of a field, as defined above, cannot be directly measured or even rigorously defined here. A field should be considered as more focused if
\begin{itemize}
\item it is structured around a limited number of important questions (or conjectures),
\item the active researchers in the field are aware of most of those questions,
\item they agree that progress on one of those questions would be highly valuable for the field.
\end{itemize}
Algebraic geometry is an example of a strongly focused field, with a number of well-known conjecture (including for instance the Riemann Hypothesis, Fermat's last theorem until it was proved in 1994, and a number of others) shaping the field. There remain a number of old, well-known and more or less central conjectures in the field.\footnote{One can consult for instance the somewhat random list on \url{https://en.wikipedia.org/wiki/List_of_conjectures} to see that Algebraic geometry, and in particular Number theory, appear prominently.}

Differential geometry was probably more focused in the 1980s than it is now, since a number of key conjectures (the Calabi and Yamabe conjecture, the Geometrization conjecture, etc) were proved between the late 1970s and the early 2000s. On the opposite, One could consider that Probability theory also became much more focused on some key conjectures from the 1980s on, thanks in particular to new connections to statistical physics ({\em e.g.} the conformal invariance of the Ising model at critical temperature) or through internal motivations ({\em e.g.} the self-intersection properties of the Brownian motion).

Other fields, such as Statistics or Numerical modelling, appear much less focused on a small number of important problems, and much more on finding new methods to solve problems that are important in applications. 

The focus of a field is related directly to its place in the awards of Fields medals, as seen in Section \ref{ssc:fields}, since Fields medals tend to be awarded to individuals who have solved (or made a key progress) in a well-known problem. The same applies to a lower extend to publications in top 5 journals, as seen in Section \ref{ssc:top3}, since the high level of selectivity of those journals means that papers that are accepted also often provide an important progress on a well-established problem. 

\begin{figure}
  \center{
    \includegraphics[width=3.8cm]{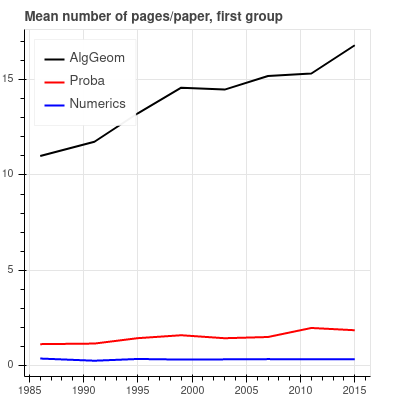}
  }
  \caption{Proportion of papers for which the word ``conjecture'' occurs in the {\em Math Reviews} entry}
  \label{fig:conjecture}
\end{figure}

One indirect way to assess the focus of a subfield of mathematics is by measuring how often the word ``conjecture'' appears in the {\em Math Reviews} entry of papers. Some results are shown in Figure \ref{fig:conjecture}. (The results are obtained by directly using the {\em mathscinet} web interface to {\em Math Reviews} and counting the number of papers in a given period with a given primary MSC code, and then the number of those for which the word ``conjecture'' appears in the entry.)

Although partial only, this data appears to confirm a relation between the focus of a field, as measured in this manner, and the status of subfields as seen above. Algebraic geometry makes a considerable use of the word, which appears in the entries of more than 15\% of articles. On the opposite, Statistics barely uses it, while Probability theory is in an intermediate situation and the use of the word is increasing markedly over the period of study.

\subsection{Departmental and institutional strategies}
\label{ssc:strategies}

The data presented in Section \ref{ssc:recruitments} provides a glimpse into the recruitments choices of mathematics departments. Can the ``status'' of subfields explain the thematic choices made by different types of departments, as seen in Section \ref{ssc:recruitments}?

Clearly, different types of departments have different needs. For instance, mathematics departments in Technology Institutes have a responsibility towards teaching mathematics to future engineers, and this could explain a stronger activity in areas like Partial differential equations or in Probability (since a strong background in probability theory is necessary for curricula in data science or quantitative finance). This might explain some specificities in Figures \ref{fig:share_hires} and \ref{fig:share_hires_rel}.

However the ``productivity'' data in Section \ref{ssc:differing} can also explain the differences in the recruitment policy of different types of departments, if we assume (as an obviously simplistic model) that departments aim at maximizing their output, in the sense of the indicator used in Section \ref{ssc:differing}. Indeed, under this hypothesis:
\begin{itemize}
\item Top departments would tend to give priority to fields with high output for top researchers, and therefore to ``high-status'' fields such as Algebraic geometry -- as seems to be the case, see Figures \ref{fig:share_hires} and \ref{fig:share_hires_rel}.
\item Lower-ranking departments might not be able to compete for the most productive researchers in the ``high-status'' fields. Figure \ref{fig:meanProd} shows that the output of authors drops quite dramatically between say the group of authors ranked 1--50 (according to their output) and those ranked 200--500. As a consequence, a better strategy for less competitive departments might be to attract highly productive researchers in a somewhat lower-status fields, such as Analysis or Algebra, or on newly fashionable fields, such as Probability. 
\end{itemize}
Clearly, more research is needed into understanding how institutions or departments choose to orient their resources towards one field or research direction or another.





\section*{Acknowledgements}

The author is grateful to Pierre-Michel Menger and Yann Renisio for many discussions, remarks and helpful comments on this work and related results, and to Fr\'ed\'erique Sachwald useful remarks on a preliminary version of the text. 

\appendix

\section{List of journals used}
\label{app:journals}

Tables \ref{tab:journals1}, \ref{tab:journals2} and \ref{tab:journals3} show the list of journals considered here, with, for each journal, its short code, the number of papers published, and the MCQ 2016 of the journal.\footnote{Errors in the table, to be corrected, in particular MCQs added not taken into account.} 


\begin{table*}[ht]
  \centering
  \begin{tiny}
  \begin{tabular}{|l|l|l|l|}
    \hline
    Journal name & Code & \# papers & MCQ 2016 \\
    \hline
    \input{listJournals1}
    \hline
  \end{tabular}    
  \end{tiny}
  \caption{List of journals considered}
  \label{tab:journals1}
\end{table*}

\begin{table}[h]
  \centering
  \begin{tiny}
  \begin{tabular}{|l|l|l|l|}
    \hline
    Journal name & Code & \# papers & MCQ 2016 \\
    \hline
    \input{listJournals2.txt}
    \hline
  \end{tabular}
  \end{tiny}
  \caption{List of journals considered, cont'd}
  \label{tab:journals2}
\end{table}

\begin{table}[h]
  \centering
  \begin{tiny}
  \begin{tabular}{|l|l|l|l|}
    \hline
    Journal name & Code & \# papers & MCQ 2016 \\
    \hline
    \input{listJournals3.txt}
    \hline
  \end{tabular}
  \end{tiny}
  \caption{List of journals considered, cont'd}
  \label{tab:journals3}
\end{table}

\bibliographystyle{amsplain}
\bibliography{/home/jean-marc/Dropbox/papiers/outils/biblio,bibsocio}

\end{document}

%% file: listJournals1.tex
ACM Trans. Math. Software & acmms & 648 & 1.72 \\ 
Acta Math. & acta & 469 & 3.51 \\ 
Acta Numer. & actaNum & 98 & 5.31 \\ 
Adv. Comput. Math. & advCompMath & 978 & 0.91 \\ 
Adv. Differential Equations & AdvDiffEq & 379 & 1.07 \\ 
Adv. Math. & advances & 4243 & 1.52 \\ 
Adv. Nonlinear Stud. & AdvNLStud & 466 & 0.88 \\ 
Adv. Theor. Math. Phys. & ATMP & 316 & 0.96 \\ 
Adv. in Appl. Probab. & AdvApplProba & 1925 & 0.64 \\ 
Adv. in Math. & advances & 208 & 1.52 \\ 
Algebr. Geom. Topol. & agt & 985 & 0.69 \\ 
Amer. J. Math. & ajm & 1595 & 1.35 \\ 
Anal. PDE & AnalPDE & 321 & 1.88 \\ 
Ann. Appl. Probab. & AnnApplProba & 1751 & 1.33 \\ 
Ann. Appl. Stat. & AnnApplStat & 647 & 0.32 \\ 
Ann. Comb. & annalsCombi & 384 & 0.61 \\ 
Ann. Inst. H. Poincar'e Anal. Non Lin'eaire & ihpan & 1250 & 2.25 \\ 
Ann. Inst. H. Poincar'e Probab. Statist. & ihpProba & 1252 & 1.18 \\ 
Ann. Probab. & AnnProba & 2881 & 1.77 \\ 
Ann. Sc. Norm. Super. Pisa Cl. Sci. (5) & pisa & 360 & 1.24 \\ 
Ann. Sci. 'Ecole Norm. Sup. (4) & asens & 846 & 2.17 \\ 
Ann. Statist. & AnnStat & 3546 & 1.65 \\ 
Ann. of Math. (2) & annals & 1547 & 3.81 \\ 
Appl. Comput. Harmon. Anal. & acha & 935 & 1.11 \\ 
Arch. Ration. Mech. Anal. & ARMA & 1015 & 2.44 \\ 
Ark. Mat. & arkiv & 692 & 0.75 \\ 
Automatica J. IFAC & Automatica & 3801 & 0.95 \\ 
Bernoulli & bernoulli & 1196 & 0.91 \\ 
Biometrika & biometrika & 2524 & 0.8 \\ 
Bull. Amer. Math. Soc. (N.S.) & bams & 974 & 0.74 \\ 
Calc. Var. Partial Differential Equations & cvpde & 500 & 1.65 \\ 
Chaos & Chaos & 1498 & 0.24 \\ 
Combin. Probab. Comput. & CombProbaComput & 1134 & 0.78 \\ 
Combinatorica & combinatorica & 1256 & 0.88 \\ 
Comm. Math. Phys. & cmp & 2578 & 1.45 \\ 
Comm. Partial Differential Equations & commpde & 2343 & 1.78 \\ 
Comm. Pure Appl. Math. & cpam & 1452 & 2.85 \\ 
Comment. Math. Helv. & cmh & 1146 & 1.03 \\ 
Compos. Math. & compositio & 940 & 1.35 \\ 
Compositio Math. & compositio & 1318 & 1.35 \\ 
Comput. Complexity & computcomp & 414 & 0.41 \\ 
Constr. Approx. & constr & 1013 & 1.0 \\ 
Discrete Comput. Geom. & dcg & 2086 & 0.61 \\ 
Discrete Contin. Dyn. Syst. & dcds & 4087 & 0.8 \\ 
Duke Math. J. & compositio & 2656 & 2.29 \\ 

%% file: listJournals2.txt
Dyn. Syst. & DynSys & 436 & 0.46 \\ 
Econometric Theory & EconTheory & 478 & 0.25 \\ 
Econometrica & econometrica & 1622 & 0.85 \\ 
Electron. Comm. Probab. & ElectrCommunProba & 908 & 0.61 \\ 
Electron. J. Combin. & elecJComb & 3070 & 0.52 \\ 
Electron. J. Probab. & ElectronJProba & 1204 & 0.99 \\ 
Ergodic Theory Dynam. Systems & etds & 949 & 0.91 \\ 
Expo. Math. & expo & 395 & 0.57 \\ 
Exposition. Math. & expo & 340 & 0.57 \\ 
Finance Stoch. & FinancStoch & 495 & 1.26 \\ 
Geom. Funct. Anal. & gafa & 1107 & 2.0 \\ 
Geom. Topol. & geotopo & 907 & 1.37 \\ 
IMA J. Numer. Anal. & imajna & 1362 & 1.5 \\ 
Indiana Univ. Math. J. & indiana & 2181 & 1.06 \\ 
Infin. Dimens. Anal. Quantum Probab. Relat. Top. & InfinDimAnal & 642 & 0.6 \\ 
Inst. Hautes 'Etudes Sci. Publ. Math. & ihes & 280 & 4.2 \\ 
Int. Math. Res. Not. & irmn & 3221 & 1.08 \\ 
Interfaces Free Bound. & interface & 221 & 0 \\ 
Invent. Math. & inventiones & 2735 & 2.89 \\ 
Inverse Problems & inverse & 3397 & 1.24 \\ 
J. Algebra & JAlgebra & 4680 & 0.62 \\ 
J. Algebraic Geom. & jag & 661 & 1.41 \\ 
J. Amer. Math. Soc. & bams & 853 & 3.56 \\ 
J. Amer. Statist. Assoc. & JAmStatAssoc & 3185 & 0.84 \\ 
J. Bus. Econom. Statist. & JBusEconStat & 355 & 0.17 \\ 
J. Combin. Theory Ser. A & JCombThA & 3072 & 1.01 \\ 
J. Combin. Theory Ser. B & JCombThB & 2012 & 1.1 \\ 
J. Comput. Graph. Statist. & JComputGraphStat & 330 & 0.4 \\ 
J. Cryptology & jcryptol & 450 & 1.47 \\ 
J. Differential Equations & jde & 6754 & 1.74 \\ 
J. Differential Geom. & jdg & 1551 & 1.56 \\ 
J. Eur. Math. Soc. (JEMS) & jems & 420 & 2.06 \\ 
J. Fluid Mech. & JFluidMech & 3467 & 0.16 \\ 
J. Funct. Anal. & cmh & 6113 & 1.31 \\ 
J. Geom. Phys. & jGeomPhys & 2843 & 0.64 \\ 
J. Lond. Math. Soc. (2) & jlms & 896 & 1.04 \\ 
J. London Math. Soc. (2) & jlms & 2328 & 1.04 \\ 
J. Math. Anal. Appl. & commpde & 20090 & 0.8 \\ 
J. Math. Biol. & JMathBio & 962 & 0.75 \\ 
J. Math. Log. & jmathlog & 84 & 1.02 \\ 
J. Math. Pures Appl. (9) & jmpa & 1326 & 1.8 \\ 
J. Mech. Phys. Solids & JMSS & 1001 & 0.42 \\ 
J. Nonlinear Sci. & jnls & 305 & 1.25 \\ 
J. R. Stat. Soc. Ser. B Stat. Methodol. & JRStatSocB & 515 & 1.83 \\ 
J. R. Stat. Soc. Ser. B. Stat. Methodol. & JRSSB & 190 & 1.67 \\ 
J. Reine Angew. Math. & crelle & 3250 & 1.3 \\ 

%% file: listJournals3.txt
J. Statist. Plann. Inference & JStatPlan & 5841 & 0.28 \\ 
J. Theoret. Biol. & JThBio & 600 & 0.22 \\ 
J. Theoret. Probab. & JTheorProba & 1442 & 0.68 \\ 
J. Topol. & jTopol & 337 & 1.15 \\ 
Math. Ann. & mathann & 4138 & 1.28 \\ 
Math. Comp. & mathcomput & 3608 & 1.3 \\ 
Math. Oper. Res. & MOR & 530 & 1.08 \\ 
Math. Program. & mathprog & 1684 & 1.38 \\ 
Math. Programming & mathprog & 1210 & 1.69 \\ 
Math. Res. Lett. & mrl & 1823 & 0.81 \\ 
Math. Z. & mathz & 4631 & 0.83 \\ 
Mem. Amer. Math. Soc. & bams & 879 & 2.5 \\ 
Nonlinearity & nonlinearity & 3222 & 1.18 \\ 
Nuclear Phys. B & NuclPhysB & 2256 & 0.14 \\ 
Numer. Linear Algebra Appl. & numLinAlgA & 1035 & 1.07 \\ 
Numer. Math. & numermath & 2720 & 1.56 \\ 
Oper. Res. & OpRe & 932 & 0.57 \\ 
Phys. D & physicad & 4908 & 0.75 \\ 
Probab. Theory Relat. Fields & PTRF & 86 & 1.81 \\ 
Probab. Theory Related Fields & PTRF & 2202 & 1.81 \\ 
Proc. Lond. Math. Soc. (3) & plms & 686 & 1.36 \\ 
Proc. London Math. Soc. (3) & crelle & 1153 & 1.26 \\ 
Publ. Mat. & PublMath & 848 & 0.78 \\ 
Q. J. Math. & qjm & 732 & 0.68 \\ 
Quart. J. Math. Oxford Ser. (2) & qjm & 549 & 0.68 \\ 
Random Structures Algorithms & random & 1206 & 1.16 \\ 
Rev. Mat. Iberoam. & rmibero & 442 & 1.11 \\ 
Rev. Mat. Iberoamericana & rmibero & 458 & 1.19 \\ 
SIAM J. Appl. Math. & siamjam & 3116 & 1.01 \\ 
SIAM J. Comput. & siamjc & 2651 & 1.24 \\ 
SIAM J. Control Optim. & siamco & 3425 & 1.08 \\ 
SIAM J. Math. Anal. & siamma & 3381 & 1.55 \\ 
SIAM J. Matrix Anal. Appl. & siamjmaa & 2160 & 1.63 \\ 
SIAM J. Numer. Anal. & siamjsc & 3766 & 1.88 \\ 
SIAM J. Optim. & siamopti & 1812 & 1.84 \\ 
SIAM J. Sci. Comput. & siamjsc & 3440 & 1.56 \\ 
SIAM Rev. & siamrev & 1165 & 1.03 \\ 
Scand. J. Stat. & ScandJStat & 447 & 0.39 \\ 
Scand. J. Statist. & ScandJStat & 838 & 0.39 \\ 
Siberian J. Differential Equations & jde & 17 & 0.0 \\ 
Statist. Sci. & StatSci & 711 & 0.54 \\ 
Statist. Sinica & StatSinica & 1646 & 0.6 \\ 
Stochastic Process. Appl. & SochProcAppl & 3435 & 1.01 \\ 
Stud. Appl. Math. & studAM & 1031 & 0.74 \\ 
Topology & topology & 1199 & 1.15 \\ 
Trans. Amer. Math. Soc. & jfa & 6956 & 1.23 \\ 